\documentclass[11pt,american]{article}
\usepackage[T1]{fontenc}
\usepackage[top=1.2in, bottom=1.2in, left=1.2in, right=1.2in]{geometry}
\usepackage{babel,amsmath,amsthm,amsfonts,graphicx,caption,subcaption,multirow,float,setspace,multirow,booktabs,collcell}
\usepackage{rotating}
\usepackage{array}
\usepackage{adjustbox}
\usepackage{lscape}

\usepackage[colorlinks=true,allcolors=blue]{hyperref}
\usepackage[flushleft]{threeparttable}
\usepackage{tikz}
\usetikzlibrary{tikzmark}
\usetikzlibrary{decorations.pathreplacing}

\newcolumntype{L}[1]{>{\raggedright\let\newline\\\arraybackslash\hspace{0pt}}m{#1}}
\newcolumntype{C}[1]{>{\centering\let\newline\\\arraybackslash\hspace{0pt}}m{#1}}
\newcolumntype{R}[1]{>{\raggedleft\let\newline\\\arraybackslash\hspace{0pt}}m{#1}}

\usepackage[super,comma,sort&compress]{natbib}
\usepackage{authblk}
\makeatletter
\newcommand{\specificthanks}[1]{\@fnsymbol{#1}}% Inserts a specific \thanks symbol
\makeatother

\begin{document}

%%%%%%%%%%%%%%%%%
\title{Zoning in American Cities: Are Reforms Making a Difference? An AI-based Analysis}
\author[1,2]{Arianna Salazar-Miranda\thanks{Corresponding Author: arianna.salazarmiranda@yale.edu}}
\author[1]{Emily Talen}
\affil[1]{University of Chicago, Chicago, IL}
\affil[2]{Yale, School of the Environment, New Haven, CT}

\maketitle
\begin{abstract}
Cities are at the forefront of addressing global sustainability challenges, particularly those exacerbated by climate change. Traditional zoning codes, which often segregate land uses, have been linked to increased vehicular dependence, urban sprawl, and social disconnection, undermining broader social and environmental sustainability objectives. This study investigates the adoption and impact of form-based codes (FBCs), which aim to promote sustainable, compact, and mixed-use urban forms as a solution to these issues. Using Natural Language Processing (NLP) techniques, we analyzed zoning documents from over 2000 U.S. census-designated places to identify linguistic patterns indicative of FBC principles. Our findings reveal widespread adoption of FBCs across the country, with notable variations within regions. FBCs are associated with higher floor-to-area ratios, narrower and more consistent street setbacks, and smaller plots. We also find that places with FBCs have improved walkability, shorter commutes, and a higher share of multi-family housing. Our findings highlight the utility of NLP for evaluating zoning codes and underscore the potential benefits of form-based zoning reforms for enhancing urban sustainability. 


\end{abstract}
\setcounter{page}{1}
\clearpage

\section{Introduction}

Cities are at the forefront of addressing global sustainability challenges, with compact urban form emerging as a key strategy to mitigate the impacts of climate change. Dense urban environments play a crucial role in reducing car dependence, which in turn lowers CO2 emissions, mitigates urban heat islands, and reduces air pollution.\cite{Acevedo2023} Compact urban environments, by prioritizing walkability, transit-oriented development, and reduced sprawl, support non-car-based mobility and protect natural habitats from encroachment. These dense, efficient urban forms also foster social connectedness and improve environmental, economic, and health outcomes, making them central to creating more sustainable, climate-resilient cities.\cite{Vallance2009, Talen2014, Sharifi2016, Duranton2020}

%cant find this citation

Yet a significant roadblock to achieving this more sustainable urban form is conventional land use zoning. Traditionally, urban planners have relied on zoning codes to dictate the configuration and development patterns of urban areas---determining what can be built, where, and how. In most places in the U.S., zoning has followed a model put in place a century ago, whereby zones were established to segregate land uses. This Euclidean, ``use-based'' approach, named after the United States Supreme Court case Euclid vs. Ambler (1926) that legalized it, focuses primarily on the separation of land uses. It has been well documented that this approach has led to a number of problems, including increased vehicular dependence, urban sprawl, and disconnection in the public realm, often undermining broader social and environmental sustainability objectives.\cite{Feitelson1993, Shen1996, Talen2013, Hsieh2019, Pendall2000, Levine2005, Knaap2007} It has been linked to segregation and displacement \cite{Manville2020, Whittemore2021, Manville2022, Shertzer2022}, and has undermined pedestrian access and social diversity. \cite{Garde2022, Talen2003, Wickersham2006} Conventional zoning is also highly restrictive \cite{Tyagi2019, Shanks2021, Song2024}, which can increase housing prices and exacerbate social separation.\cite{Gyourko2011, mleczko_2023}

In light of these problems, calls have been made to reform zoning to be more supportive of sustainable urban form, specifically allowing compact, mixed-use, pedestrian-oriented development.\cite{Ghorbanian2020} Such codes would be more ``form-based'' than ``use-based,'' involving such reforms as: ending single-family only zoning, allowing a wide mix of housing types (including accessory dwelling units or ADUs, which are thought to be inherently affordable), eliminating or reducing parking requirements, and having much lower minimum requirements for lot size, dwelling unit size, and building setbacks.\cite{Gray2022, LiangLinlin2024} Since the emphasis is on form rather than use, land uses would be more mixed. These reforms would translate to much less restriction, whereby variation of sizes, tenures, types and building ages would be permitted \cite{jacobs_1961, Talen2012}, which in turn would allow a mix of rents and prices and thus enable social diversity.

Researchers have begun to explore the implications of these reforms. Most of this research has been qualitative, for example by investigating the link between FBCs and conventional zoning for individual cities\cite{Garde2017}, or between FBCs and urban design quality\cite{Hansen2014, Chin2024}, developer decision-making processes\cite{Hughen2017}, and sustainability.\cite{Garde2017, Garde20172, Zhang2022} FBCs have also been evaluated indirectly by looking at the relationship between urban design quality of the kind that FBCs promote and outcomes like walking behavior and mental health\cite{Ameli2015, Sung2015, Buttazzoni2023} There is some evidence that neighborhoods regulated by FBCs have held their value and even increased in worth during the 2008 economic downturn\cite{EnvironmentalProtectionAgency2011}, and that FBC census tracts were correlated with lower median rent and higher percent rental.\cite{Talen2021} Other studies have linked FBCs to higher residential turnover and displacement\cite{Park2017, Tagtachian2019}, and faulted FBCs for failing to address the issues of vacancy and lack of affordable housing.\cite{DenoonStevens2020, ozay2022} But in an extensive literature review, Borys and Benfield (2019) compiled the results of 135 studies that include some form of measurement of the impact of FBCs and concluded that the compact, mixed-use development of FBCs have substantial benefits for people (in terms of, e.g., physical well-being, affordability, social capital), the planet (e.g., reduction of greenhouse gas emissions, land conservation), and profit (e.g., jobs per acre, property value).\cite{HazelBorys2019} The studies they reviewed focused on small-scale effects associated with individual jurisdictions.

Despite the importance of zoning reform and the increasing popularity of FBCs, large scale quantitative evaluation is rare. The analysis is hampered by the fact that the text of zoning codes is unstructured and highly variable, which makes it difficult to assess codes quantitatively and comparatively across different regions. A major challenge is identifying FBCs in the first place, as municipalities may adopt FBC elements---such as mixed use and narrower setback requirements---in ways that are not explicitly labeled as FBC in their codes. This subtlety makes it difficult to systematically identify and document FBC adoption using traditional methods, such as through surveys, which are time consuming and often produce low response rates.\cite{Bronin2023} 
%One recent study showed that planners ``systematically'' underestimate the restrictiveness of local zoning codes.\cite{oneill2024} 
Attempts to collect zoning information in a more systematic way are ongoing, but efforts such as the National Zoning Atlas rely on manual input of zoning information.\cite{Sahn2021}

We address these issues by employing Natural Language Processing (NLP) to assess FBC-related zoning in more than 2000 census-designated places across the U.S. NLP offers a promising approach to identifying the linguistic patterns and terms that signal the adoption of FBC principles, even when not explicitly stated.  Moreover, NLP allows for large volumes of text in zoning documents to be analyzed at scale, allowing us to conduct a comprehensive and systematic assessment of FBC adoption and its impacts. Researchers have begun to employ NLP and other algorithmic approaches to evaluate zoning in a more scalable way\cite{mleczko_2023, Song2024}, but our focus is different in that we focus exclusively on zoning reform---i.e., form-based codes.

Our contributions are two-fold. First, we provide a method to delineate zoning codes that are more ``form-based'' than ``use-based'', which proxies as a way to determine the existence of an FBC. As previously noted, zoning codes do not fall neatly into  ``reformed'' or ``form-based'' vs. ``conventional'' or "use-based" categories, but are instead a matter of degrees. To delineate FBC codes, we construct a metric of language similarity that reveals how closely the language of zoning aligns with FBC principles. This offers a novel perspective on the extent of FBC influence, revealing underlying linguistic similarities that may be unrecognized by municipalities themselves. The measure reveals that FBC adoption is widespread throughout the country, with slightly higher adoption in the South, often in larger municipalities, and does not vary significantly across demographic characteristics, such as average income, population, and education. This extends previous research (e.g., Talen, 2021) that relied on incomplete lists of explicitly adopted form-based codes. We also show that our FBC measure captures aspects related to the character of development, thus differing from other research employing NLP methods to assess zoning restrictiveness.

Second, we investigate linkages between FBCs and urban form, using both built environment measures and urban form proxies based on walkability and commuting distances. Our findings are largely consistent with previous qualitative literature. Specifically, we find that FBCs are associated with higher floor-to-area ratios, narrower and more consistent street setbacks, and smaller plot sizes---three dimensions associated with sustainable urban form. In terms of urban mobility and housing metrics, we find that the adoption of FBCs is associated with higher walkability scores, shorter commute distances, and with an increase in the share of multi-family housing, although the results for housing are less precise than the others. Additionally, we demonstrate that these associations are evident for neighborhoods developed after 1950. This provides some support for the view that the estimates reflect the impacts of previous zoning reforms rather than a reverse causality story in which places with better ex-ante morphological attributes were more likely to adopt zoning codes with FBC elements. 

\section{Results}\label{sec:results}

We obtained zoning codes from Municode, the largest repository of municipal codes in the U.S, using a custom-built web scraper in 2023. Municipal codes encompass a wide variety of documents, including ordinances, resolutions, and regulations that govern different aspects of municipal governance, such as zoning ordinances, building codes, environmental regulations, and public safety rules. The structure and classification of these documents can vary significantly from one municipality to another. For instance, zoning regulations might be explicitly titled as ``Zoning'' in one municipality, while in another, they might be categorized under ``Streets, Sidewalks, and Public Property.'' 

To ensure we focused on zoning-related content, we filtered the chapters to include only those with titles explicitly related to zoning. We applied filters to identify chapters where key phrases appeared within the first few lines of text. These phrases included “Zoning”, “Land division”, “Subdivision”, “Land use”, “Land development”, “Streets”, “Master Plan”, “Development”, and “Neighborhood Design”. This approach helped us accurately capture zoning-related texts across municipalities with diverse terminologies and document structures.

Our sample comprises 2,723 municipalities and designated areas, each matched to its corresponding census place, which serves as our unit of analysis (see section \ref{sec:method} for sample selection details).

\subsection*{Delineating FBCs}

To determine which codes in our sample are more vs. less in line with FBC principles, we first constructed a set of FBC criteria extracted from 70 codes that have been explicitly identified as FBCs. These codes came from two sources: a list maintained by Placemakers.com, a planning and design firm that operates in the U.S. and Canada, and a list of award-winning FBCs maintained by the Form Based Codes Institute. With these official FBCs as our reference, we could then measure the extent to which zoning codes in our Municode sample align with FBC principles. 

To do this, we segmented each zoning code into smaller, manageable chunks, resulting in an average of 12 chunks per document. Each chunk was then processed using the BigBird model to generate semantic embeddings, which are numerical representations that capture the meaning of the text. We averaged these embeddings to create a single embedding for each document, representing its overall semantic content. Summary statistics, including the total tokens and average tokens per chunk, are presented in Table \ref{table: document statistics} and Figure \ref{fig:summary}.

Next, we compared these document embeddings to the reference set of 70 FBCs by calculating a cosine similarity score. The cosine similarity score measures the similarity between two vectors (in this case, the embeddings), based on the cosine of the angle between them. This measure is useful to understand how closely each municipal zoning code aligns with the principles embodied in the FBC documents. A higher score indicates a stronger thematic alignment with FBC principles.

\subsubsection*{Thematic Differences Between FBCs and Traditional Zoning}

We explore thematic differences between FBC codes and other types of zoning documents in three ways. First, we implemented Principal Component Analysis (PCA) on the document embeddings. PCA is a dimensionality reduction technique that transforms high-dimensional data into a lower-dimensional space, capturing the most variance with fewer dimensions. This method allows us to visualize the primary thematic patterns by projecting the documents onto the first two principal components.

\begin{figure}[!ht]
    \begin{center}
    \includegraphics[width=\textwidth]{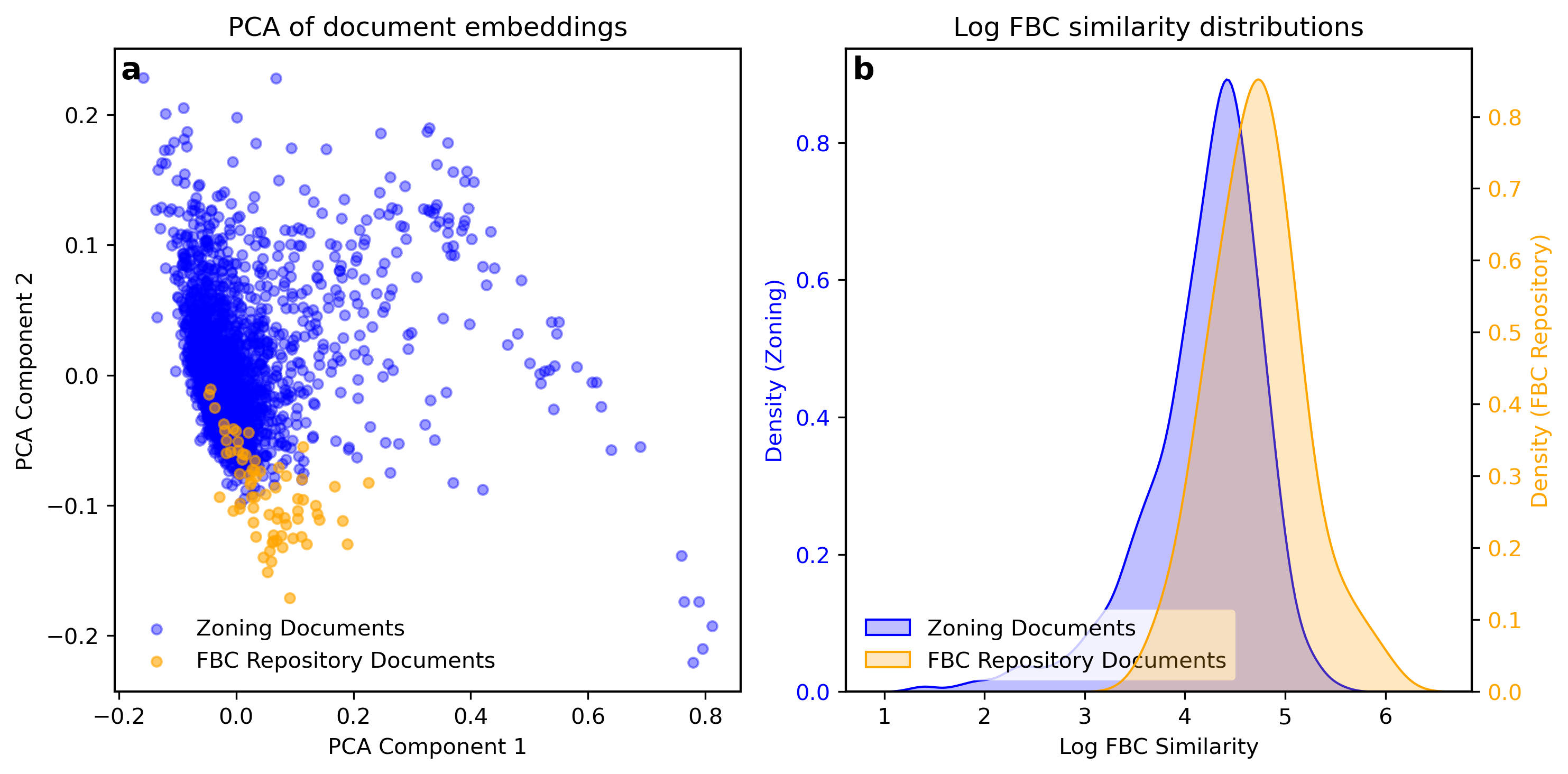}
    \end{center}
    \caption{\textsc{Differences in thematic content.} Panel (a) PCA on document embeddings, showing separate embeddings for the reference set and for FBC codes obtained from our two repositories. Panel (b) FBC similarity distributions for zoning documents (N=2,723) and reference set documents (N=70). }
    \label{fig:pca}
\end{figure}

Figure \ref{fig:pca} shows a scatter plot of the zoning documents projected onto the first two principal components. The figure shows that FBC documents form a tight cluster, distinct from the broader spread of other zoning documents. This tight clustering indicates that FBC documents have strong internal thematic coherence, implying that these documents are guided by a consistent set of themes. In contrast, the dispersion among other zoning documents suggests a more varied thematic base, reflecting the diversity in traditional zoning practices.

In our second approach, we examined the main themes across three groups of documents: (1) 50 random documents from the top 20\% of the FBC distribution, (2) 50 random documents from the bottom 80\% of the FBC distribution, and (3) all FBC repository codes explicitly identified as form-based. From these texts, we extracted all paragraphs mentioning setbacks or FAR, using a predefined list of terms to account for the fact that these themes might be referenced differently across places (see Appendix for details).

To analyze these paragraphs, we used ChatGPT-4 to extract recurring themes and terms. Specifically, for each document, we prompted the model to provide an overview of the main points related to setbacks or FAR and to identify the most frequently occurring themes and terms associated with these topics. These outputs captured recurring patterns in the zoning content, which we then analyzed using Term Frequency-Inverse Document Frequency (TF-IDF) vectorization. TF-IDF assigns a numerical score to each term, reflecting how frequently it appears in a group relative to its overall frequency in the corpus. This allowed us to highlight terms that are both frequent and distinctive within each group, which we interpret as themes because they capture recurring and distinctive patterns of meaning in the zoning content.

\begin{figure}[!ht]
    \begin{center}
    \includegraphics[width=\textwidth]{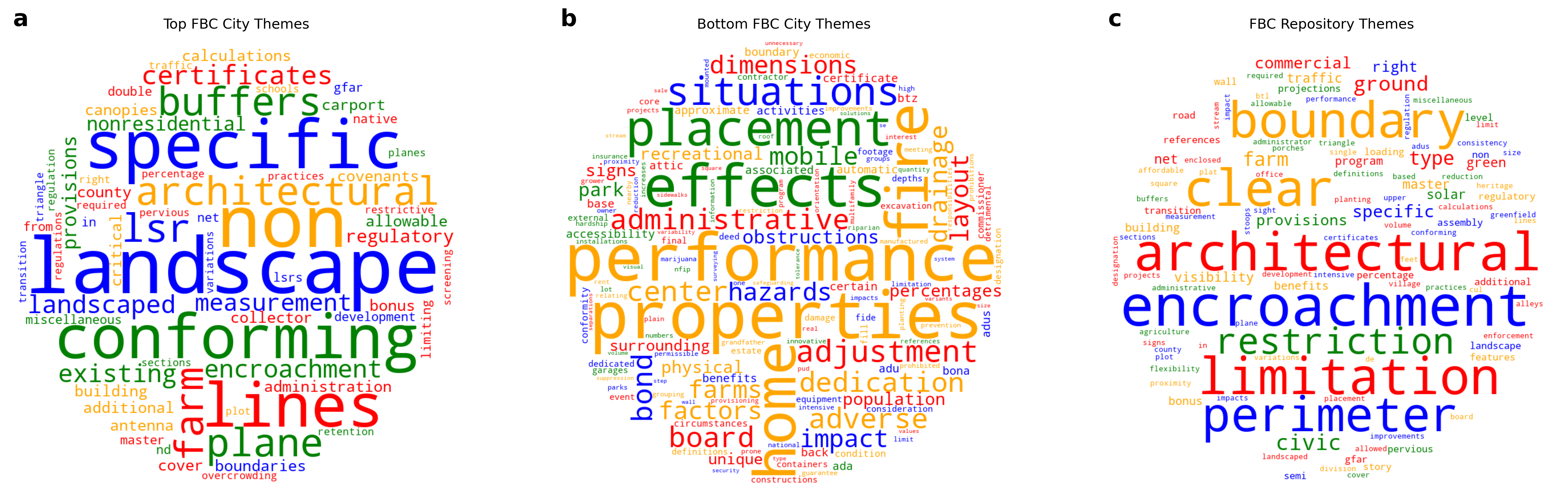}
    \end{center}
    \caption{\textsc{Word clouds for places with top and bottom FBC.} The figure illustrates the prominent themes extracted from zoning documents of top and bottom cities. Panel (a) represents themes prevalent in top FBC cities (top 20\% of FBC distribution), panel (b) shows the themes from bottom FBC cities (bottom 80\% of FBC distribution), and panel (c) shows the themes for the FBC repository documents. Themes were extracted using the ChatGPT-4 natural language model. A TF-IDF vectorizer was used to identify significant themes, and the difference in TF-IDF scores between top and bottom cities was used to filter the themes displayed. The size of each word reflects its distinctiveness within the group, as determined by TF-IDF, which accounts for both frequency and uniqueness relative to the entire corpus. The colors are included for visual clarity and do not carry analytical significance. Common words were excluded to highlight unique themes in each category. See section \ref{sec:method} for a full description of the prompts used to obtain the themes from the documents.}
    \label{fig:wordclouds}
\end{figure}

Figure \ref{fig:wordclouds} shows word clouds for each group. Word size reflects TF-IDF scores, representing the distinctiveness of a term within its group relative to the overall corpus.  Documents with the highest FBC scores (panel a) prominently feature terms such as ``landscape'', ``architectural'', ``buffers'', and ``conforming''.  These terms reflect fundamental FBC principles, emphasizing physical urban form and place-making concepts, such as design standards, landscape requirements, and prioritization of architectural elements. Slightly different terms but a similar focus on place and urban form are reflected in the word cloud for the FBC repository themes (panel c).

In contrast, documents with the lowest FBC scores (panel b) highlight terms such as ``performance'', ``adverse'', ``placement'', and ``hazards''. These terms suggest a focus on the legal and administrative aspects of zoning, such as property restrictions, hazard mitigation, and performance standards.

Finally, we compare FBC vs. traditional zoning using a commonly used measure of regulatory stringency: the Wharton Residential Land Use Regulatory Index (WRLURI). Figure 3 shows a scatter plot between FBC similarity and the WRLURI, revealing the distribution of places across different combinations of regulatory stringency and FBC alignment.  Each point represents a place, color-coded based on its quadrant classification.

\begin{figure}[!ht]
    \begin{center}
    \includegraphics[width=0.8\textwidth]{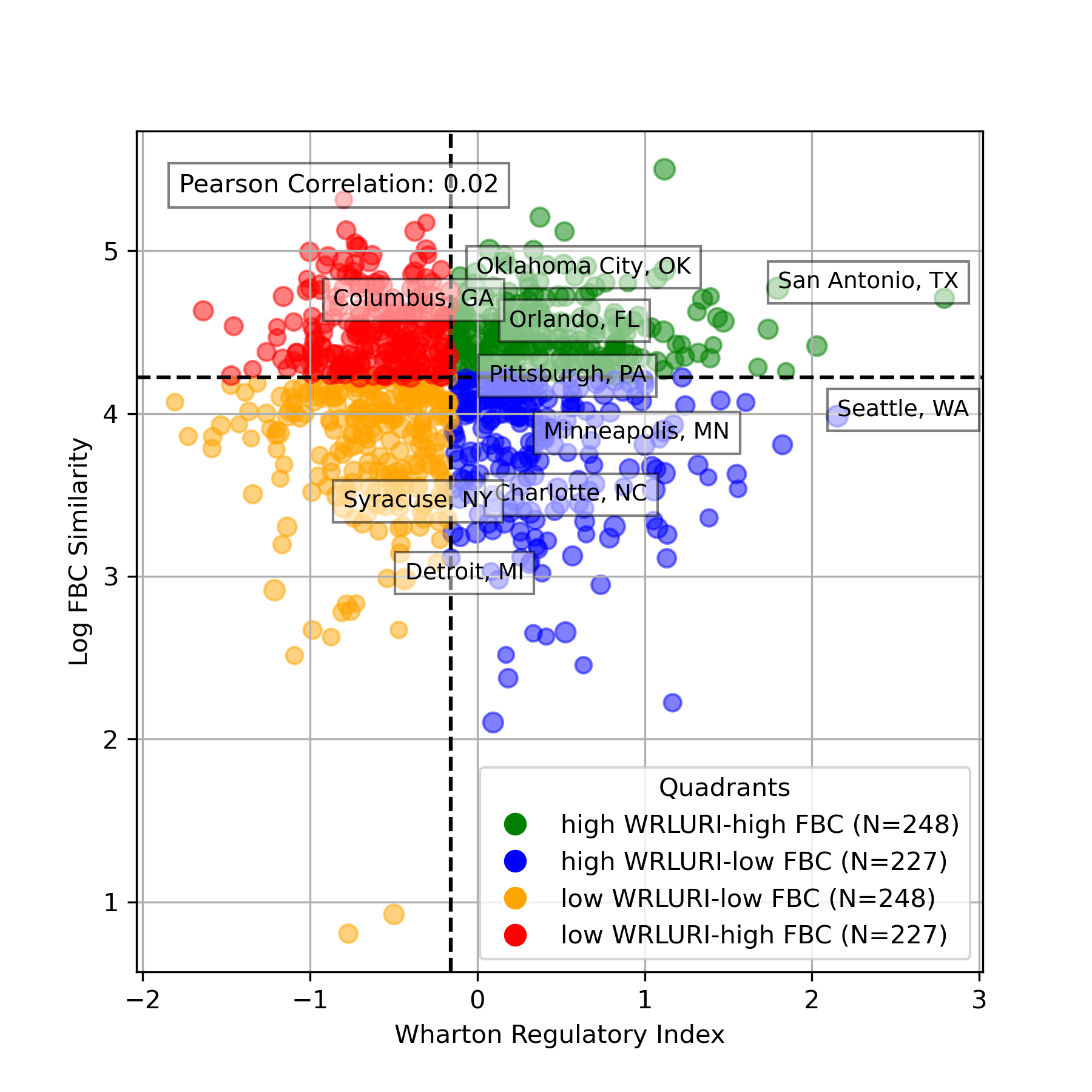}
    \end{center}
    \caption{\textsc{Relationship between FBC similarity and Wharton Regulatory Index (WRLURI).} The figure is divided into four quadrants based on the median values of WRLURI and log FBC similarity. Each point represents a place, with colors indicating the quadrant: high WRLURI-high FBC (green), high WRLURI-low FBC (blue), low WRLURI-low FBC (orange), and low WRLURI-high FBC (red). Dashed lines represent the median values dividing the quadrants. Cities with high populations within each quadrant are labeled. The Pearson correlation coefficient between WRLURI and log FBC similarity is 0.02, indicating a weak relationship. Quadrant counts are as follows: high WRLURI-high FBC (N=248), high WRLURI-low FBC (N=227), low WRLURI-low FBC (N=248), and low WRLURI-high FBC (N=227).}
    \label{fig:quadrants}
\end{figure}

Figure \ref{fig:quadrants} shows that FBC similarity and regulatory restrictiveness are uncorrelated (Pearson correlation of 0.02). The figure also shows a wide variation in FBC similarity across places with similar levels of regulatory restrictiveness. In other words, zoning codes can be highly restrictive and form-based, or they can be highly restrictive and not form-based. For instance, Charlotte and Orlando have comparable regulatory restrictiveness, yet Orlando exhibits a much higher FBC similarity than Charlotte. This pattern is not uncommon. The figure highlights that many places conform to FBC principles but have varying levels of regulatory restrictiveness. For example, Orlando and Oklahoma City exhibit high FBC similarity despite high regulatory restrictiveness. Conversely, cities like Columbus show high FBC similarity but low regulatory restrictiveness. Cities such as Seattle and Minneapolis exhibit high regulatory restrictiveness but low FBC similarity, while Syracuse has both low regulatory restrictiveness and low FBC similarity.

Overall, the lack of relationship and the wide variation between FBC similarity and regulatory restrictiveness suggest that adopting FBC is not necessarily about increasing zoning stringency. The wide variation in FBC similarity among places with comparable regulatory stringency suggests that FBC adoption is driven by goals other than merely increasing restrictions. Some cities may use FBCs to enhance aesthetic or urban design outcomes, while others may prioritize density, even at similar regulatory levels. The weak relationship between FBC similarity and regulatory restrictiveness might also suggest that form-based codes are being applied in a variable way, from highly restrictive to a more permissive application, depending on the city's overall planning goals. Overall, adopting FBCs does not automatically signal a shift toward stricter land use control.

\subsubsection*{Geographic and Demographic Differences Among FBCs}

\begin{figure}[!ht]
    \begin{center}
    \includegraphics[width=\textwidth]{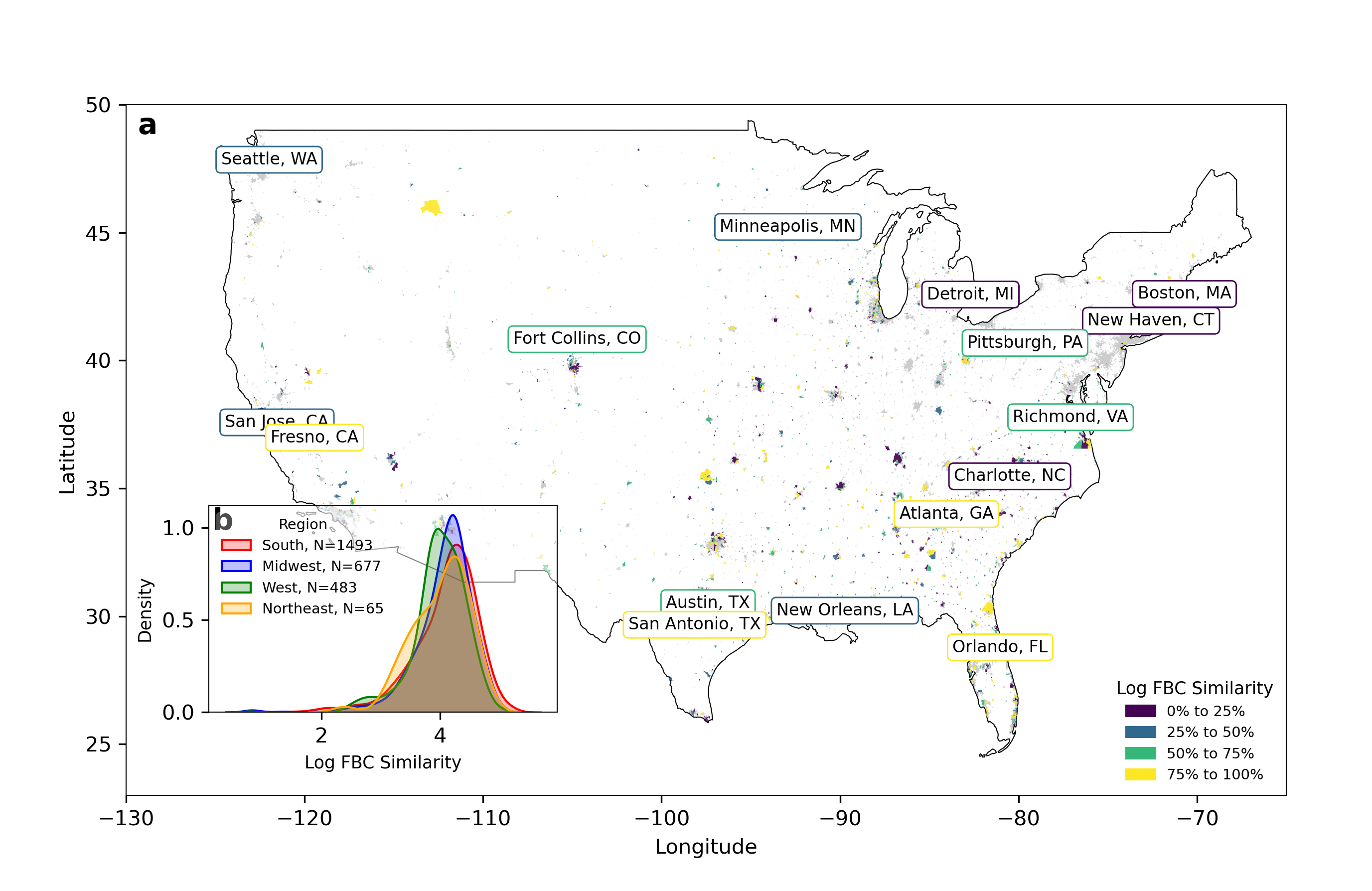}
    \end{center}
    \caption{\textsc{FBC similarity in the United States.} Panel (a) shows FBC similarity for all census places in the United States (N=2,723), highlighting 3 cities with high population in each FBC similarity quantile. Panel (b) plots the distribution of FBC similarity by region. 
}
    \label{fig:map}
\end{figure}

Using our FBC similarity measure, we explore the FBC code distribution across the U.S., both in terms of geography and demographics. Figure \ref{fig:map}, panel (a) maps the geographic distribution of FBC similarity by quantiles, highlighting four cities in each quantile. The map shows that zoning codes with high FBC are widespread across the country, with slightly higher values in Southern cities. One potential explanation for the slightly higher adoption of FBC in Southern cities is the strong presence of New Urbanism in the region. For example, the popularity of New Urbanism in Florida, a planning movement strongly aligned with FBCs, has likely contributed to the broader acceptance and implementation of FBC principles in local zoning practices. 

Figure \ref{fig:map}, panel (b) presents the distribution of FBC similarity by region, showing that most of its influence occurs \textit{within} regions rather than \textit{between} them. The figure illustrates that within each region, there is a broad range of FBC scores, indicating diverse levels of FBC influence among places. A variance decomposition analysis reinforces this finding, showing that differences in FBC influence between regions account for less than 1\% of the total variation in FBC similarity scores. In contrast, variations within regions make up 99\% of the total variation in FBC scores. This suggests that local factors, rather than regional characteristics, play a more significant role in determining the extent of FBC adoption.

We also explore whether places that adopted FBCs have a different demographic composition. This is important to understand if FBC codes are exclusionary or are predominant in whiter and richer areas. Our data shows that this is not the case.  Table \ref{table: summary statistics} reports the mean and standard deviation of a number of demographic characteristics, showing that places with high and low FBC similarity are comparable in terms of race, income, and other demographics. 

	\begin{table}[!ht]
		\centering
		\caption{\sc{Summary Statistics}}
		\label{table: summary statistics}
		\resizebox{1\textwidth}{!}{\begin{tabular}{L{7cm}C{3cm}C{3cm}C{3cm}}\toprule\toprule
				&\multicolumn{3}{c}{\sc{Samples}}\\\cmidrule(r){2-4}
				\vspace{0.2cm}
				&\multicolumn{1}{c}{All Sample} & 	\multicolumn{1}{c}{High FBC} & 	\multicolumn{1}{c}{Low FBC} \\
				&\multicolumn{1}{c}{(N=2,723)} & 	\multicolumn{1}{c}{(N=544)} & 	\multicolumn{1}{c}{(N=2,179)} \\
				\\
				&\multicolumn{3}{c}{\sc{Panel I. Demographics}}\\\cmidrule(r){2-4}
				\multirow{2}{7cm}{Population (Log)\dotfill}&    9.198 &     9.538 &     9.114 \\  
				& [    1.494] &  [    1.358] &  [    1.514) \\ 
				\multirow{2}{7cm}{Median Income\dotfill}&61205.336 & 58220.344 & 61950.549 \\  
				& [31708.662] &  [25088.665] &  [33118.905) \\ 
				\multirow{2}{7cm}{\% College Degree\dotfill}&   28.728 &    28.389 &    28.813 \\  
				& [   16.806] &  [   14.484] &  [   17.339) \\ 
				\multirow{2}{7cm}{\% Foreign Born\dotfill}&    9.484 &     9.413 &     9.502 \\  
				& [   10.361] &  [    9.775] &  [   10.505) \\ 			
				\multirow{2}{7cm}{\% Over 65 Years\dotfill}&   17.186 &    16.619 &    17.328 \\  	
				& [    7.685] &  [    7.003] &  [    7.841) \\ 
				\multirow{2}{7cm}{\% Owner Occupied Housing\dotfill}&   62.888 &    61.294 &    63.286 \\  	
				& [   14.242] &  [   13.423] &  [   14.415) \\ 
				\multirow{2}{7cm}{\% White\dotfill}&   75.026 &    73.906 &    75.306 \\  
				& [   20.505] &  [   19.751] &  [   20.684) \\ 
				\multirow{2}{7cm}{Unemployment Rate\dotfill}&    5.711 &     5.561 &     5.748 \\  	
				& [    3.391] &  [    3.025] &  [    3.476) \\    	
				\\\bottomrule
		\end{tabular}}
		\begin{minipage}{1\linewidth}											
			\textsl{Note.---Columns 1 to 4 report the sample means and standard deviation (in square brackets) for the demographic variables in each row. Column 1 reports summary statistics for the full sample. Column 2 reports summary statistics for neighborhoods with high FBC (defined as the top 20\% places according to this metric). Column 3 reports summary statistics for neighborhoods with low FBC (defined as the bottom 80\% places according to this metric). }								
		\end{minipage}	
	\end{table}

\subsection*{Linking FBCs to Urban Form}

Thus far we have presented a method of FBC delineation and showed how FBCs differ from traditional zoning codes in terms of thematic and geographic dimensions. Our second central question concerns the linkages between FBCs and urban form characteristics.  While zoning regulations are designed to shape urban development by controlling land use, building heights, densities, and setbacks, the actual impact on the built environment and the resultant effect on people’s behavior is uncertain. Here, we investigate some possible associations.

We first create three urban form measures that capture important aspects of the built environment: setbacks, floor area ratios (FARs), and minimum plot sizes. Proponents of FBCs have emphasized their potential to increase density (sometimes measured as FAR) and reduce setbacks and plots sizes, aiming to create built environments with better street-to-height proportions and consistent retail frontage. These dimensions are crucial because they can enhance walkability, decrease car dependency, and support local businesses by creating a more engaging and pedestrian-oriented public realm.

\subsubsection*{Setbacks, FARs, and plot sizes}

We construct our morphological measures of urban form using detailed geospatial data on the location and dimension of plots and buildings, including their height, obtained from LandGrid and Microsoft Building Footprints. Specifically, we compute four variables: FAR, median setbacks, the setback standard deviation, and minimum plot size. For setbacks, we first calculate the average setback at the street level using street geometries obtained from OpenStreetMap. This involves averaging the setbacks of all buildings along each street segment. Next, we aggregate these street-level averages to compute the overall average setback for the entire place, ensuring that our final measure reflects the typical street setback across the place. 

For the FAR measure, each building is spatially joined with its corresponding land use plot. The FAR for each building is calculated as the ratio of the total built area to the plot area, and we average the FAR values at the place level. The minimum plot size variable refers to the smallest plot size within each place. Further details on these variables can be found in section \ref{sec:method}. 

We compare our measures to two types of places: those in the top 20\% for FBC scores, termed ``high FBC,'' and the remaining 80\%, termed ``low FBC.'' Results using the continuous measure are comparable and can be found on Table \ref{table: urban form continuous} in the Appendix. 

Using these measures, we estimate regression models of the form:

\begin{equation} \label{eq:1}
\textrm{Urban Form Outcome}_{i, r, v} = \alpha_{r} + \beta\cdot\textrm{FBC}{i, r, v} + \delta_{i} + \theta\cdot X_{i,r} + \epsilon_{i,r,v}
\end{equation} 

For each outcome, $i$ denotes a place, and we estimate equation \eqref{eq:1} at the place level. The regression explains the urban form for a place located in region $r$ as a function of the following factors: $FBC_{i, r}$, which captures the degree of alignment with FBC principles. A positive coefficient $\beta$ would suggest that places with higher FBC similarity scores exhibit distinct urban form characteristics. We also include zoning vintage fixed effects $\delta_{i}$ to account for the time that zoning has been in place. $X_{i,r}$ is a vector including place characteristics and geographic covariates that account for differences that might influence urban forms, such as differences in regional economic conditions or cultural attitudes towards urban form, and includes covariates such as latitude and longitude and place characteristics, such as its type (borough, city, town, village) and size (area). $\epsilon_{i,r,v}$ is the error term.

\begin{figure}[!ht]
    \begin{center}
    \includegraphics[width=\textwidth]{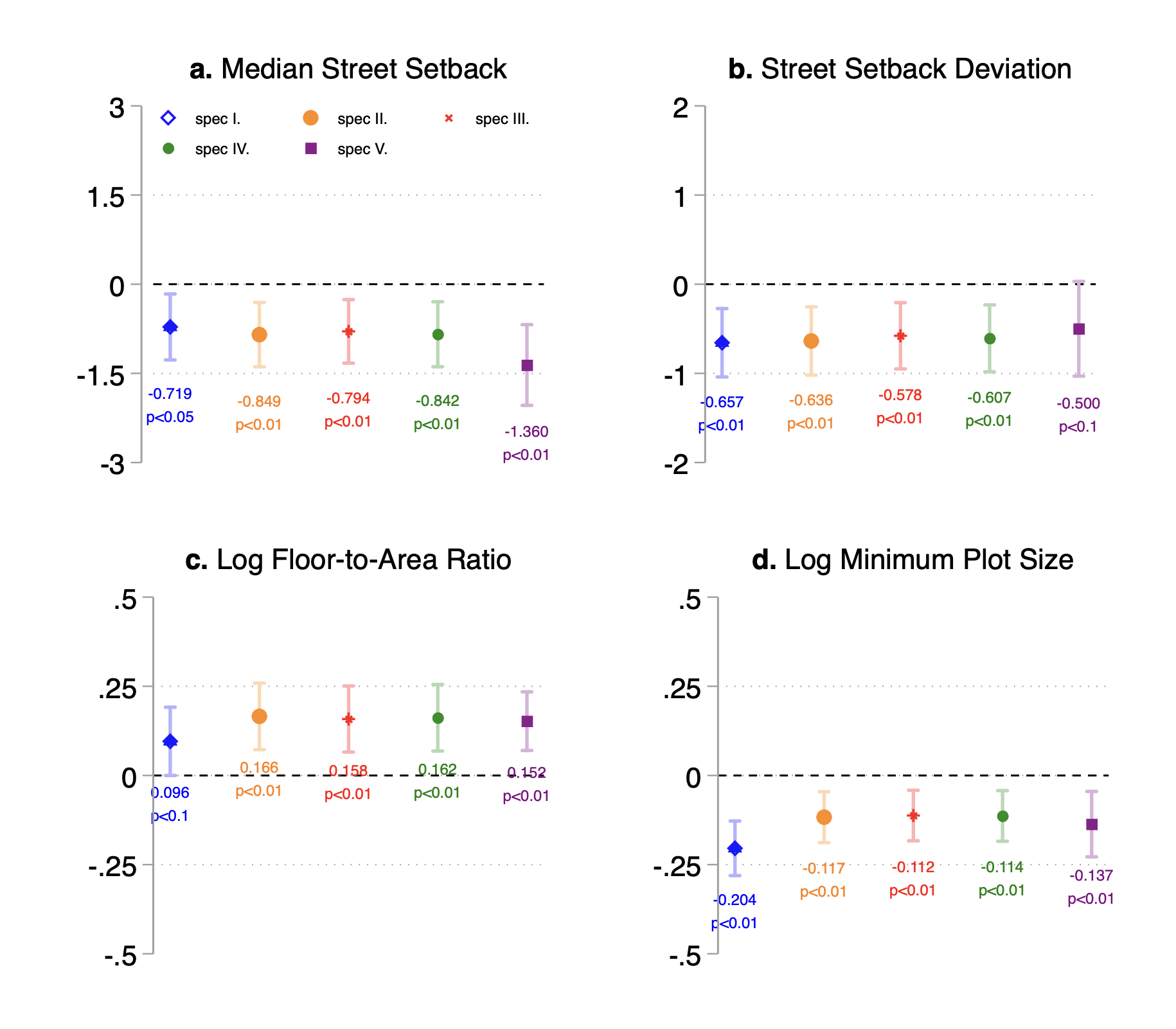}
    \end{center}
    \caption{\textsc{OLS estimates of high FBC and morphological measures of urban form.} The figure plots point estimates from regressing urban form outcomes on high-FBC (top 20\% of the FBC distribution), using five different specifications. The unit of observation is the census place. Panel (a) plots estimates for median street setbacks. Panel (b) plots estimates for street setback deviation. Panel (c) shows the estimates for log floor-to-area ratio (FAR). Panel (d) plots estimates for minimum plot size. Each figure plots the point estimate from these regressions with 95\% confidence intervals (point estimate ± 1.96 * SE). Specification I includes state fixed effects, Specification II adds location controls (latitude-longitude), and Specification III further controls for log area (km²) and type of place (borough, city, town, village). Specification IV incorporates zoning vintage dummies (1982-1996, 1996-2008, 2008-2016, 2016-2021). Specification V reports estimates for outcomes computed for neighborhoods developed after 1950 in each place. The number of observations for each outcome is 2,452 for median street setbacks and street setback deviations, 2,450 for log FAR, and 2,452 for log minimum plot size. See Table \ref{table: urban form discrete} for point estimates and confidence intervals. 
}
    \label{fig:reg1}
\end{figure}

Figure \ref{fig:reg1} plots the point estimates from equation \eqref{eq:1} along with 95\% confidence intervals for these three outcomes (the top panel of Table \ref{table: urban form discrete} reports the same estimates and standard errors). Panel (a) reports the estimates for median street setbacks as the dependent variable, Panel (b) for street setback deviation, Panel (c) for floor-to-area ratio (FAR), and Panel (d) for minimum plot size. The first specification includes state-fixed effects and location controls, specifically latitude and longitude. The second and third specifications further control for log area (km$^2$) and type of place (borough, city, town, village), respectively. The fourth specification incorporates zoning vintage dummies. The fifth is analogous to the fourth specification but reports estimates for outcomes computed for neighborhoods developed after 1950 in each place. 

Figure \ref{fig:reg1} presents four key findings related to the relationship between FBC and urban form. First, across all specifications, we observe that high FBC places exhibit significantly lower street setbacks compared to low FBC places. The negative coefficients in column four of Panel (a) imply that moving a place from the bottom 80\% of the FBC distribution to the top 20\% is associated with narrow setbacks of 0.84 meters on average. This finding suggests that FBC is associated with reduced street setbacks, aligning with the goal of creating more pedestrian-friendly environments with buildings closer to the street.

Panel (b) shows that high FBC places also have significantly lower street setback deviations. The negative coefficient in column four implies that moving a place from the bottom 80\% of the FBC distribution to the top 20\% is associated with a reduction in street setback deviations of 0.60 meters. This reduction in setback variability indicates a more consistent urban form, which is a key objective of FBC---to ensure uniformity in street frontage and urban design.

Panel (c) indicates that high FBC places have a higher FAR. Specifically, the positive coefficient in column four implies that moving a place from the bottom 80\% of the FBC distribution to the top 20\% is associated with an FAR increase of 16\%. This result suggests that FBC is associated with increased density, a key objective of FBCs designed to promote walkable communities. 

Finally, Panel (d) turns to minimum plot size and shows that high FBC places have smaller plot sizes. In particular, estimates in column 4 show that moving a place from the bottom 80\% of the FBC distribution to the top 20\% is associated with a decrease in minimum plot size of 11\%. 

Taken together, our results support the view that FBCs have a stronger relationship with denser, more human-scaled built environments, with buildings closer to the street, than zoning codes that are less form-based.

\subsubsection*{Walkscore, commute distance, and housing}

We now explore the association between FBCs and three urban mobility and housing measures: walkscore, commute distance, and the share of multi-family housing. We use these measures to re-estimate equation \eqref{eq:1} (the bottom panel of Table \ref{table: urban form discrete} reports the same estimates and standard errors). 

\begin{figure}[!ht]
    \begin{center}
    \includegraphics[width=\textwidth]{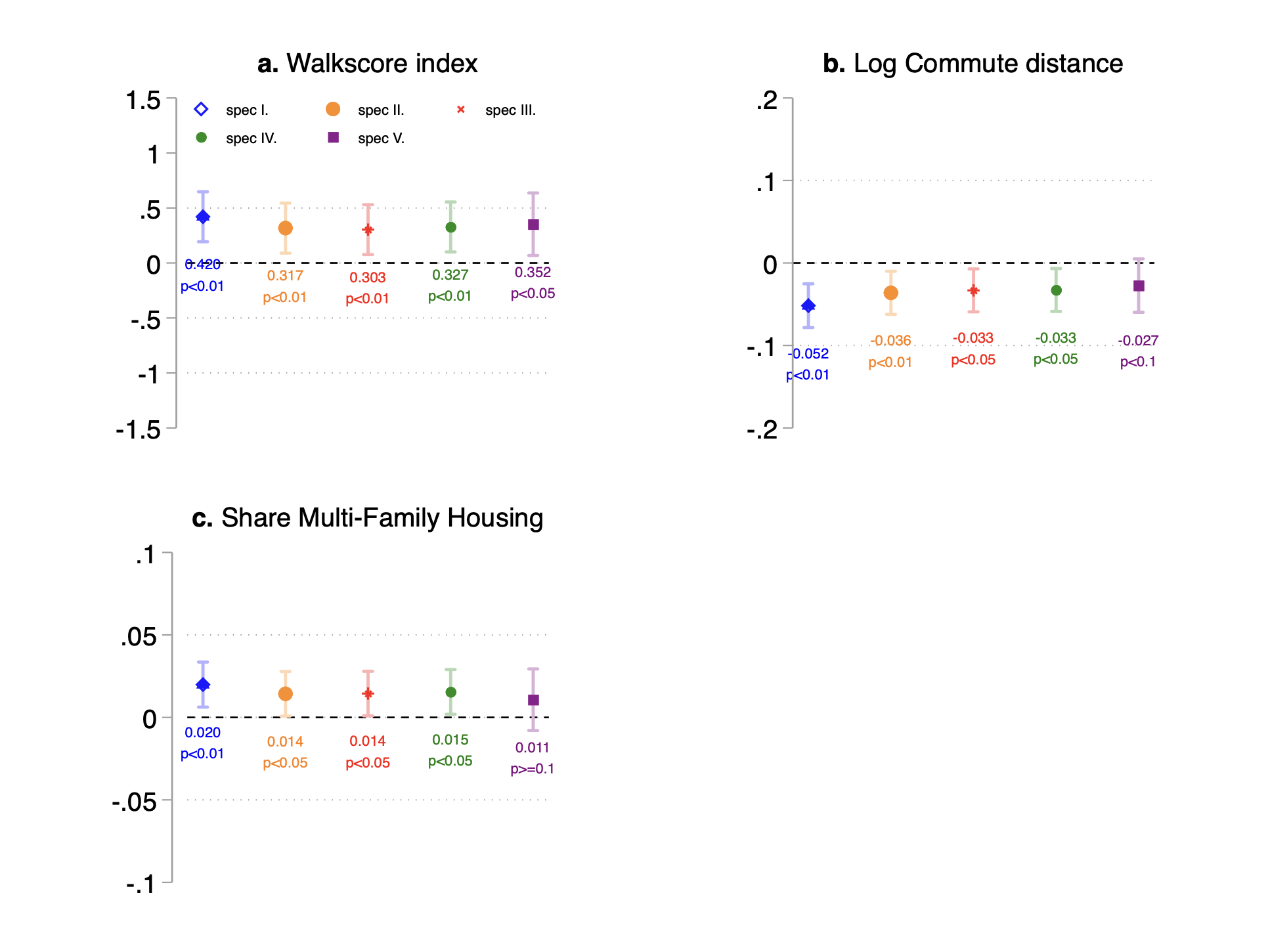}
    \end{center}
    \caption{\textsc{OLS estimates of high FBC and urban mobility and housing measures of urban form.} The figure plots point estimates from regressing behavioral outcomes on high-FBC (top 20\% of the FBC distribution), using four different specifications. The unit of observation is the census place. Panel a plots estimates for walkscore. Panel b plots estimates for commute distance (km). Panel c shows the estimates for the share of trips within 15 minutes. Each figure plots the point estimate from these regressions with 95\% confidence intervals (point estimate ± 1.96 * SE). Specification I includes state fixed effects, Specification II adds location controls (latitude-longitude), Specification III further controls for log area (km²) and type of place (borough, city, town, village), and Specification IV incorporates zoning vintage dummies (1982-1996, 1996-2008, 2008-2016, 2016-2021). Specification V reports estimates for outcomes computed for neighborhoods developed after 1950 in each place. The number of observations for walkscore, commute distance, and multi-family housing is 2,687, 2,566, and 2,718, respectively. See Table \ref{table: urban form discrete} for point estimates and confidence intervals. 
}
    \label{fig:reg2}
\end{figure}

Panel (a) in Figure \ref{fig:reg2} shows that places with higher FBC have higher walkscores. The estimates in column four imply that moving a place from the bottom 80\% of the FBC distribution to the top 20\% is associated with a walkscore increase of 0.32. Given that the average walkscore is 8.52, this change represents an increase of 3.76\% relative to the average walkscore, a rather modest, though nevertheless positive, effect. 

In Panel (b), we examine the association between FBC similarity and commute distance. The estimates show that high FBC places have shorter commutes. In particular, moving a place from the bottom 80\% of the FBC distribution to the top 20\% is associated with a 3.3\% decrease in commute distance. This finding indicates that FBC principles are associated with improved walkability and also appear to correlate with shorter commuting patterns in a significant way. 

Panel (c) plots the association between FBC similarity and the share of multi-family housing. The results reveal a positive effect of FBC similarity on the share of multi-family housing. Specifically, moving from the lower to the upper quintile of the FBC distribution results in a 1.5\% increase in the share of multi-family housing. 

Taken together, these results suggest that FBCs are associated with higher walkability and proportion of multi-family housing. However, the results on housing become less precise when using the continuous FBC measure as an explanatory variable (see Table \ref{table: urban form continuous} in the Appendix).

\subsubsection*{Disentangling the role of FBCs}

Our results describe correlations between areas with stronger FBC influence and their built environment. One interpretation is that these estimates reflect the causal impact of adopting FBC codes on urban form, aligning with the belief that FBC codes can enhance sustainability and promote denser, more walkable environments.

However, this is only one of several plausible interpretations. First, our estimates could reflect selection bias: areas with initially superior morphological attributes (i.e., more compact and dense) may have been more likely to adopt zoning codes conforming to FBC principles. In this scenario, our estimates would reflect reverse causality rather than the causal impact of FBCs. Second, our estimates might be confounded by omitted variables. For example, whiter and wealthier places might inherently be more walkable and dense, to begin with, and may also be more likely to adopt FBC codes.

Although we cannot entirely rule out these alternatives, we present two pieces of suggestive evidence that cast doubt on them. First, we exploit the timing of neighborhood development. In particular, we estimate the same regression models presented in Section \ref{sec:results}, but focus only on neighborhoods developed after 1950 in each place. Since most zoning reforms were adopted in the 1920s and 1930s, these codes preceded the development of these neighborhoods, alleviating concerns about reverse causality.

Results from this analysis are shown in specification 5 of Figures \ref{fig:reg1} and \ref{fig:reg2}. Across all specifications, we observe similar point estimates of FBC's impact on post-1950 outcomes. 
%These findings lend support to the notion that our estimates may indeed reflect the causal effect of zoning on the built environment.

Second, Table \ref{table: summary statistics} reports summary statistics for high and low FBC places, showing no significant differences in their demographic makeup. This suggests that our estimates are not confounded by observable demographic differences, although it does not rule out the possibility of unobserved factors influencing both FBC adoption and positive urban form outcomes.

\section{Discussion}\label{sec:discussion}

In the quest for sustainable cities, many planners advocate for a new approach to development regulation, with form-based codes (FBCs) emerging as a dominant strategy. Prioritizing form rather than use, as well as increasing allowance for density, mixed land use, mixed housing types, and smaller unit and lot sizes, FBCs have been at the forefront of zoning reform over the past two decades. But reforms are sometimes successive and a matter of gradation, infiltrating zoning codes incrementally, such as through setback decreases or allowances for mixed-use and higher density in some locations. Because these reforms are layered and embedded in codes in a variety of ways, natural language processing (NLP) was leveraged to filter through our large set of municipal zoning documents to gauge FBC similarity. We were then able to characterize the contexts in which FBCs are more prevalent by measuring their association with certain kinds of urban forms. Previous research relating FBCs to urban form has mostly relied on a more qualitative assessment.

We applied several techniques to illustrate what our measure of FBC similarity was capturing. First, despite the nuances and complexities of zoning code changes and reforms (form-based codes are not always explicitly labeled as such), we found that FBC qualities comprised a discernible and distinct type of zoning code, distinguishable from other types of codes. Documents with high FBC characteristics were tightly clustered in the PCA scatter plot, indicating strong internal thematic coherence. Traditional zoning documents, on the other hand, were more dispersed and had a broader thematic diversity. While there was some overlap between FBC codes and traditional zoning documents, true FBC documents showed a narrow distribution of high similarity scores, another indication of their strong thematic consistency.

Second, this internal thematic consistency also aligned with what FBCs are trying to achieve, which is a less car-dependent, more pedestrian-oriented urban quality. Codes with high FBC similarity featured themes related to physical urban form and place-making (e.g., ``landscape,'' ``architectural''), while low FBC similarity documents were more focused on the legal and administrative aspects of zoning (e.g., ``performance,'' ``hazards''). This qualitative difference is exactly what FBC proponents have been aiming for \cite{HazelBorys2019}, giving more attention to the built environment---the form of a city--- and less attention to rules and procedures that could result in unpredictable outcomes, including car-based urban form.\cite{Wickersham2006, Gyourko2011, Tyagi2019, Ghorbanian2020, Shanks2021, Song2024, mleczko_2023} An administrative focus equates with a neglect of place quality, and is exactly what traditional zoning codes are critiqued for.\cite{Hansen2014}

One important consideration when using large language models to evaluate zoning documents is whether the resulting classifications align with expert human assessments. We conducted a validation exercise to assess agreement in our specific context. We selected ten zoning codes, each from the top and bottom of the FBC similarity distribution, and extracted all paragraphs related to setbacks and FAR. We then developed a rubric that evaluated how well the documents reflected key principles of FBCs, including their emphasis on design and urban form, support for active and mixed-use streetscapes, and context-sensitive approaches to regulation. A human evaluator, blinded to the FBC similarity scores, applied this rubric, as detailed in the Supplementary materials. The results showed a reasonable degree of alignment: for 17 out of the 20 locations, the evaluator's assessments were consistent with the FBC scores.

We were able to make two other conclusions about FBCs: that they are not linked to zoning restrictiveness and that high FBC similarity scores are distributed across the US, where variations within regions are more significant than variations between regions. The first conclusion is interesting in that there is a popular perception that FBCs are attempting to be more rather than less restrictive.\cite{perez2016formbasedcodes} Our data suggests the opposite is the case. The second conclusion is interesting in that it suggests that local factors play an important role in FBC adoption. This points to a related conclusion: that local effects and governance play a significant role in shaping urban sustainability, particularly in relation to urban form, and these local effects might very well outweigh national or state-level provisions, which is in line with recent calls for a ``New Urban Agenda'' based on local governance.\cite{satterthwaite2016newurbanagenda, silva2020localgovernance} The findings, in other words, highlight the importance of deepening our understanding of local processes and the need to build capacity at the local level to support sustainability-focused decision-making.

Our second main task was to characterize the built environment associated with high vs. low FBC similarity. In the case of places with high FBC similarity, we found that there was a significant association between code and outcome for three essential urban form characteristics: setbacks, floor area ratios, and plot size.  As compared to low FBC places, high FBC places had shorter setbacks, higher floor area ratios, and smaller plot sizes, all of which, as discussed above \cite{Hansen2014}, support the goal of creating walkable, pedestrian-friendly environments. These positive associations extended to more urban mobility and housing measures: walkability (as measured by walkscore), commute time, and multi-family housing, in line with other FBC research.\cite{Talen2012, Hughen2017, Garde2017, HazelBorys2019, Zhang2022}. Also consistent with previous studies, our findings suggest that FBCs are strongly associated with creating denser and more human-scaled environments.\cite{Ghorbanian2020, Talen2021} 

Our analysis is exploratory and focuses on revealing correlation rather than causality, the approach is still valuable as a way of identifying \textit{potential} causal relationships. FBCs seem to be enjoying an exponential rise, so it is important to be able to gauge whether FBC associations are in the expected and desired direction. If they were not – i.e., if FBCs had been shown to correlate with lower walkscores, longer commutes, less multi-family housing, larger setbacks—it would be cause for concern. FBC proponents should find comfort in our analysis as it supports the expected direction of what FBCs are trying to accomplish.

Analyzing the association between code and built form is important because if the code differs significantly from the built environment it is meant to effect, this signals either that the code is being tasked with making significant hoped---for changes to place quality, or that the code is not having the intended effect. In either case, the zoning code is in a weakened position, and regulation of the private market, to which zoning is directed, might not have the ability to produce a more sustainable urban outcome. In that case, more public sector involvement via capital investment and direct development subsidy may offer greater hope for producing the desired change.

While our methodology is unique, there are several limitations that should be noted. One limitation of ChatGPT is that the model makes sense of foreground data (input) by leveraging its background data (training); the reliance on past data without real-time learning limits the accuracy and applicability of its responses. Additionally, the use of  TF-IDF for theme quantification emphasizes term frequency, potentially overlooking less frequent but contextually significant themes and failing to capture semantic relationships between terms. The overall process depends heavily on the quality of input. We also note that certain explanatory metrics that we use, such as walkscore, can be measured in alternative ways.

Future research could seek to refine the methodology used in this study, perhaps by using human coders to review and refine the summaries generated, and to ensure that important nuances, domain-specific knowledge, and context are captured. Alternative methods of relating the content could be explored, such as the use of topic modeling to reveal patterns in the data and making the insights from the content more explicit and easier to interpret. Further differentiation of FBC places based on urban intensity levels, and how that relates to urban form outcomes, could be explored. Additionally, more research is needed to determine the appropriate time period for evaluating FBC outcomes. Having data on the timing of coding reform is difficult to track, but doing so would allow causal exploration.

While form-based coding draws from past practices, it is still a relatively new concept with largely unknown effects. Planners are justified in their optimism about zoning reform, but it is crucial to be realistic about the capabilities and limitations of FBCs in addressing significant issues facing U.S. cities. There is a widespread belief that cities need to be less wasteful, more efficient, less land-consumptive, more compact, less car-dependent, and more equitable. Planners view zoning reform as a key driver of this change. To meet these expectations, FBCs should undergo regular evaluation, using the tools of NLP to aid in this process.

\clearpage
\section*{Methods}\label{sec:method}

The methods section is organized into two primary sections. In the ‘Datasets’ section, we explain the datasets used in our analysis. In the ‘Data processing’ section, we detail the data processing procedures that we followed to construct the document embeddings and the FBC similarity measure. 

\subsection*{Datasets}

\noindent\textbf{Zoning Codes}\\
\indent Our primary source of municipal documents is the \href{https://library.municode.com/}{Municode database}, the largest repository of municipal codes in the U.S. We developed a custom web scraper to download all available chapters for each municipality, saving each chapter as a text file. 
 
Additionally, we obtained ground truth data from the \href{formbasedcodes.org}{Form-based codes Repository}. This repository gathers exemplary FBC codes from various communities across the U.S. These documents serve as benchmarks for identifying FBC principles in municipal codes. From this repository, we selected 70 codes that pertain to entire cities or neighborhoods, as opposed to smaller projects or specific developments. Of these 70 codes, 32 are Driehaus Prize winners. The Driehaus Prize recognizes communities with exemplary FBC zoning codes.\\

It should be noted that one limitation of Municode is that it is a pay-to-use service and thus might exclude smaller and less affluent jurisdictions. In addition, there may be a lag between when a new ordinance is adopted and when it appears in the Municode database. 

Municode does not provide finer spatial granularity that indicates specific neighborhoods or sub-areas within the municipality. As a result, we are unable to determine where an FBC applies only to a subset of the jurisdiction, which effectively means we assume that what is written in the zoning code applies to the entire municipality. \\

\noindent\textbf{US Census}\\
\indent We extracted all the demographic and housing characteristics from the 2010 Decennial Census at the place level. Places represent closely settled communities identified by name, such as municipalities, cities, towns, villages, boroughs, and communities, among others. \\ 

\noindent\textbf{Development year of neighborhoods}\\
\indent For the exercise where we exclude neighborhoods developed after 1950, we compute all outcomes at the census block group level. We determine the development year of each neighborhood using data from Leyk \& Uhl (2018), which provides pixel-level values at a spatial resolution of 250 meters, showing the year in which the first structure was established in that area, with a temporal granularity of five years.\cite{Leyk2018}

The property information is obtained from the ZTRAX dataset produced by the Zillow Group, encompassing details from around 200 million parcels.

To establish the founding year of each neighborhood, we use the modal year of building construction within that neighborhood. This may not provide an exact age of the neighborhood due to possible building renewals and redevelopment. However, it serves as a proxy for the neighborhood's actual age. The building stock ages can change due to renewals and redevelopments. Therefore, we consider the development year as either the original year of construction or the most recent year in which a neighborhood experienced significant renewal or redevelopment.\\

\noindent\textbf{Zoning vintage dummies}\\
To control for the vintage of zoning codes in our regression analysis, we extracted the date listed in each zoning document. We started by storing the first 500 words of each document, as this section typically includes the dates. We then used the GPT-4 model to analyze this text and return a table with columns for the year the code was adopted, the year it was originally published, the year it was republished, and the earliest year mentioned.

The following prompt was used to extract zoning dates via ChatGPT:
‘The following text contains a zoning document. Return a table with the following columns: (1) Year the code was adopted/enacted, (2) Year the code was originally published, (3) Year the code was republished, and (4) Earliest year mentioned in the document. Separate each column value with a "\$" symbol for easy parsing.'
To construct the zoning vintage variable, we prioritized the year the code was adopted (N=1605). If the adoption year was missing, we used the year it was originally published (N=1108). The remaining 1,108 were filled using the original published year. For the remaining documents  (N=5), we attempted to use the earliest year mentioned in the document, but none of these documents had that information available. This resulted in a final sample of 2,718 documents with non-missing years. 

To validate the results, we randomly sampled 20 documents and confirmed that the extracted dates were accurate.

To construct the zoning vintage variable, we prioritized the year the code was adopted. If the adoption year was missing, we used the year it was originally published. If both of these were missing, we used the earliest year mentioned in the document. \\

\noindent\textbf{LODES}\\
\indent Commute length data was obtained from the 2019 Longitudinal Employer-Household Dynamics Origin-Destination Employment Statistics (LODES). This dataset provides aggregate job counts linked to home and work census blocks. We computed the average commute distance as the mean Haversine distance from work to home, weighted by the total jobs in each census block group and then aggregated across places.\\

\noindent\textbf{Land Use and Building Geometry}\\
\indent Parcel data was obtained from Regrid data by LOVELAND Technologies, collected in 2023. Building footprints were obtained from the Microsoft Maps building footprint data, which provides information on building areas and heights. These measures are derived from computer vision algorithms applied to high-resolution satellite imagery. Microsoft reports a false positive ratio of less than 1\% in a sample of 1000 randomly selected buildings, indicating high data quality.\cite{Microsoft} \\

\noindent\textbf{Roads}\\
\indent Street data was obtained from OpenStreetMap using the OSMNX API for each place.\cite{Boeing2024} Street data is in shapefile format, providing precise road locations. We used road geometries to compute setback distances from property lines and building lines to the nearest street, further explained in the data processing section. \\

\noindent\textbf{Wharton Regulatory Index}\\
\indent We used the Wharton Residential Land Use Regulation Index (WRLURI) to assess the stringency of local regulatory environments. WRLURI is derived from a comprehensive survey of residential land use regulations conducted across over 2,600 U.S. communities.\cite{Gyourko2011} The index captures various aspects of the land regulatory process, including the complexity and length of project approval processes, costs associated with lot development, and the extent of regulatory barriers. WRLURI is widely recognized for providing a standardized measure of regulatory environments, facilitating comparisons across different jurisdictions.

35\% of these communities overlap with our sample, and therefore, our analysis comparing the FBC measure with WRLURI is limited to a subset of 950 places. \\

\textbf{Walkscore}\\
\indent We used the walkscore measure produced by the Environmental Protection Agency (EPA) to evaluate walkability. The EPA’s walkscore index predicts walkability based on features in the built environment that promote walking. The index considers factors such as land use diversity, street connectivity, and ease of access to public transit. Each census block is categorized into four groups and scored from 1 (least walkable) to 20 (most walkable). More details on the methodology can be found in the EPA’s technical documentation.

The measure of walkscore is constructed as the average walkscore for each place, weighted by the total population in each block group.\\

\subsection*{Data Processing}

\noindent\textbf{Sample construction}\\
\indent The goal of our sample construction was to select chapters specifically related to zoning across comparable geographies. The challenge was twofold: First, municipal codes use varied terminology for zoning, with some codes referring to "zoning" while others use terms like "land division" or "land development." Second, municipal codes include a broad range of topics, and many chapters are unrelated to zoning.

To address this, we used the fact that each chapter in Municode is saved as an independent file with a clear title and chapter number. We developed a script to automatically scan the first five lines of each chapter for zoning-related terms such as "Zoning," "Municipal Code," "Land division," "Subdivision," "Land use," "Land development," "Streets," "Master Plan," "Development," and "Neighborhood Design." The script checked for both uppercase and lowercase variations of these terms. If any of these terms appeared within the first five lines, the chapter was flagged as zoning-related and included in our sample.

After identifying relevant chapters, we compiled the documents at the municipality level and matched them to either census place, county, or county subdivision. In cases where a municipality name matched multiple census geographies, we prioritized place matches over county and county subdivision matches, and county subdivision matches over county matches unless 'county' was explicitly mentioned in the name.

To maintain a consistent geographic unit, we restricted our sample to census places, which encompass municipalities and designated areas, comprising 87\% of the Municode documents, resulting in a sample of 2,721 places.
\\

\noindent\textbf{Embedding creation}\\
\indent We preprocessed the documents for semantic analysis by converting texts to lowercase and removing non-alphabetic characters and stopwords. We used the NLTK library for stopword removal and created a custom list to include legal and geographic terms common across all codes.

To analyze the zoning codes, we used the BigBirdModel, specifically chosen for its ability to process long documents, which is particularly suited for the extensive nature of zoning codes. We implemented the model with pre-trained weights from Hugging Face's Transformers library. We removed the classification layer and reconfigured it to generate semantic embeddings rather than perform classification tasks.

Given the extensive length of zoning documents, we segmented them into chunks of up to 4096 tokens (the maximum allowed by the model) with no overlap. Each chunk was then processed through the BigBird model to generate semantic embeddings. These embeddings were then averaged across chunks for each document to produce a single, representative vector, capturing the overarching semantic features of the text. The embeddings for each document were then normalized to ensure that each document’s representation was comparable.

The averaging process ensures that each document, regardless of its length or number of chunks, is represented by a single embedding vector of the same dimensionality. While longer documents generate more chunks, the averaging step normalizes their influence so that the final embedding captures the overall semantic structure rather than just length. Table \ref{table: document statistics} and Figure \ref{fig:summary} present summary statistics of document length.

We segmented the text into non-overlapping chunks to manage the extensive length of zoning documents and the computational demands. While this approach may separate some logically connected content, zoning codes are generally structured and technical, making them less reliant on continuous narrative flow. Therefore, we anticipate that the overall semantic representation remains largely preserved. \\

\noindent\textbf{FBC similarity measure}\\
\indent To evaluate the thematic alignment of zoning documents with established FBCs, we constructed a cosine similarity score. This score measures the cosine of the angle between two vectors (or documents) in a multi-dimensional space, providing a measure of semantic proximity between documents. The cosine similarity score was calculated between the embeddings of each municipal zoning document in our dataset and the centroid of the FBC documents, capturing how closely each municipal zoning code aligns with the principles embodied in the FBC documents. 

To analyze the thematic content of zoning documents, we used Principal Component Analysis (PCA) on the document embeddings. PCA was used to reduce the high-dimensional space of the document embeddings into two principal components that capture the most significant variance in the data. This allowed us to visualize the thematic differences between FBC codes and other zoning documents.\\

\noindent\textbf{FAR measures}\\
\indent Our FAR measures are computed at the parcel level. We use the Microsoft footprint data that provides us with information on the height and footprint of each building and combine it with parcel data from LOVELAND, which provides the area of the parcels. These data are obtained from local governmental agencies, state agencies, and third-party hosts. 

The total built area for each building was calculated by multiplying the building's area by its height. This provided a measure of the total built area for each building. In cases where building height data was missing or marked as invalid, we imputed these values using the average building height within the same zoning district. Each building was spatially joined with its corresponding land use plot to determine the plot area. The FAR for each building was calculated as the ratio of the total built area to the plot area. We then averaged the FAR values at the place level. \\

\noindent\textbf{Setbacks Measure}\\
\indent To compute the setbacks measures we use spatial data for data for buildings, land use plots, and street networks. 

For each building, we calculated the centroid and decomposed the building into its sides. We then computed the distance from the building side that faced the nearest street and recorded this distance. Additionally, we calculated the distance from the plot boundaries to the nearest street. The setback deviation was computed by subtracting the plot's setback distance from the building's setback distance. To obtain a measure of building placements along a street, we averaged the setbacks and setback deviations for each street segment. Finally, we took the median of these averages across all places to derive a single measure of setbacks for each place in our sample.

To mitigate the effects of drawing errors in defining parcels and errors in classifying building boundaries, we implemented several data adjustments. First, we addressed extreme values in setback deviations by winsorizing the top 1\% of these values, thereby reducing the influence of outliers that could skew the analysis. Second, for buildings identified as being located outside their plot boundaries, which likely resulted from drawing inaccuracies, we assigned a building setback value of zero.\\

\noindent\textbf{Thematic Content}\\
\indent To explore the thematic content captured by our FBC similarity measure, we analyzed zoning documents with high and low FBC similarity scores and the explicitly identified documents from the FBC repository. Our process involved several steps. 

First, we selected 50 random documents from the top 20\% of the FBC distribution and 50 random documents from the bottom 80\% of the FBC distribution. This stratified sampling allowed us to compare the thematic content of documents with high and low alignment to FBC principles and to evaluate how these alignments compare to the principles explicitly reflected in the FBC repository.

Next, we extracted paragraphs that specifically mentioned setbacks and floor-to-area ratio (FAR). To capture a broad range of relevant content, we used predefined lists of terms related to these themes. For setbacks, the terms included ``setback,'' ``building line,'' ``frontage requirement,'' ``yard requirement,'' ``buffer zone,'' ``setback distance,'' ``architectural clear zone,'' ``perimeter restriction,'' ``boundary requirement,'' ``zoning setback,'' and ``encroachment limitation.'' For FAR, the terms included ``FAR,'' ``floor area ratio,'' ``floor space ratio,'' ``floor space,'' ``site ratio,'' ``buildable area,'' ``building coverage,'' ``development intensity,'' ``development density,'' and ``intensity of land use.'' This ensured that we included all relevant discussions on these themes, even if they were referred to differently across various documents.

We used ChatGPT-4 to analyze the extracted paragraphs. For each document, we asked the model to ``Provide a brief overview of the main points related to setbacks discussed in the document.'' The exact prompt used was: ``Provide a brief overview of the main points related to [setbacks terms] discussed in the following text: [text].'' Following this, we asked the model to ``Identify the most frequently appearing themes and terms related to setbacks.'' The exact prompt used was: ``Identify the most frequently appearing themes and terms related to [setbacks terms] in the following text: [text].'' The output from these prompts provided a structured summary of the text, including recurring terms and concepts, which we interpreted as thematic content.

To evaluate the consistency and coherence of these themes across documents, we took a sample of 10 summaries from the top 20\% of FBC similarity scores and 10 summaries from the bottom 80\%. We concatenated these summaries within each group to create composite documents representing the dominant thematic patterns in each category.

We then applied Term Frequency-Inverse Document Frequency (TF-IDF) vectorization to the composite documents to quantify the importance of each identified term. This step highlights terms that are not only frequent within a given group but also distinctive relative to the other groups. TF-IDF allows us to identify the terms that best capture the thematic distinctions between the high-FBC group, the low-FBC group, and the FBC repository documents.

\noindent\textbf{Validating the FBC Similarity Measure with Human Coders}\\
\indent 

To evaluate whether the classifications from the large language model align with human assessments, we conducted a validation exercise focused on setbacks and FAR---two elements central to zoning.

We selected 10 zoning codes each from the top and bottom of the FBC similarity distribution, representing those most and least similar to FBCs. For each zoning code, we extracted all paragraphs that referenced setbacks and FAR (see Thematic Content for details on extraction methods).

A human evaluator reviewed the extracted text for each zoning document and scored it according to rubrics for setbacks and FAR. The rubrics included three dimensions for each zoning element:

\textbf{Setbacks:}
\begin{enumerate}
\item Connects setbacks to the design of public spaces or streetscapes.
\item Uses setbacks to enhance walkability or the pedestrian experience.
\item Mentions setbacks in the context of urban form.
\end{enumerate}

\textbf{FAR:}
\begin{enumerate}
\item Emphasizes urban design and form over functionality.
\item Supports mixed-use and/or active streetscapes.
\item Contextually regulates density and intensity instead of applying uniform standards.
\end{enumerate}

For each zoning document, the evaluator assigned a score of 1 to any dimension addressed in the document. The evaluator was blinded to the FBC similarity scores to avoid bias.

We compared the human-assigned classifications with the LLM-derived FBC similarity scores. For 17 out of the 20 zoning codes, the human classifications aligned with the FBC similarity-based classifications, demonstrating a reasonable degree of alignment. 

Tables \ref{table: setbacks} and \ref{table: far} present the validation results, showing the rubric scores and associated descriptions for each zoning document.

\newpage
\bibliographystyle{unsrt}
\bibliography{fbc}

\begin{thebibliography}{10}

\bibitem{Acevedo2023}
A~Acevedo-De-los R{\'{i}}os.
\newblock {Building Materials and the Climate: Constructing a New Future}.
\newblock Technical Report No. UCL-Universit{\'{e}} Catholique de Louvain, UCL-Universit{\'{e}} Catholique de Louvain, 2023.

\bibitem{Vallance2009}
Suzanne Vallance, Harvey~C Perkins, and Jennifer~E Dixon.
\newblock {Compact Cities: Everyday Life, Governance and the Built Environment An Annotated Bibliography and Literature Review}.
\newblock Technical report, University of Auckland, 2009.

\bibitem{Talen2014}
Emily Talen and Julia Koschinsky.
\newblock {Compact, walkable, diverse neighborhoods: Assessing effects on residents}.
\newblock {\em Housing Policy Debate}, 24(4):717--750, 2014.

\bibitem{Sharifi2016}
Ayyoob Sharifi.
\newblock {From Garden City to Eco-urbanism: The quest for sustainable neighborhood development}.
\newblock {\em Sustainable Cities and Society}, 20:1--16, jan 2016.

\bibitem{Duranton2020}
Gilles Duranton and Diego Puga.
\newblock {The economics of urban density}.
\newblock {\em Journal of Economic Perspectives}, 34(3):3--26, 2020.

\bibitem{Feitelson1993}
Eran Feitelson.
\newblock {The Spatial Effects of Land Use Regulations A Missing Link in Growth Control Evaluations}.
\newblock {\em Journal of the American Planning Association}, 59(4):461--472, 1993.

\bibitem{Shen1996}
Q~Shen.
\newblock {Spatial Impacts of Locally Enacted Growth Controls: The San Francisco Bay Region in the 1980s}.
\newblock {\em Environment and Planning B: Planning and Design}, 23(1):61--91, 1996.

\bibitem{Talen2013}
Emily Talen.
\newblock {Zoning For and Against Sprawl: The Case for Form-Based Codes}.
\newblock {\em Journal of Urban Design}, 18(2):175--200, 2013.

\bibitem{Hsieh2019}
Chang-Tai Hsieh and Enrico Moretti.
\newblock {Housing Constraints and Spatial Misallocation}.
\newblock {\em American Economic Journal: Macroeconomics}, 11(2):1--39, apr 2019.

\bibitem{Pendall2000}
Rolf Pendall.
\newblock {Local Land Use Regulation and the Chain of Exclusion}.
\newblock {\em Journal of the American Planning Association}, 66(2):125--142, 2000.

\bibitem{Levine2005}
Jonathan Levine.
\newblock {\em {Zoned out: Regulation, markets, and choices in transportation and metropolitan land-use}}.
\newblock Resources for the Future, 2005.

\bibitem{Knaap2007}
Gerrit-Jan Knaap, Huibert~A Hacco{\^{u}}, Kelly~J Clifton, and John~W Frece, editors.
\newblock {\em {Incentives, Regulations and Plans}}.
\newblock Number 4039 in Books. Edward Elgar Publishing, dec 2007.

\bibitem{Manville2020}
Paavo~Monkkonen {Michael Manville} and Michael Lens.
\newblock {It's Time to End Single-Family Zoning}.
\newblock {\em Journal of the American Planning Association}, 86(1):106--112, 2020.

\bibitem{Whittemore2021}
Andrew~H Whittemore.
\newblock {Exclusionary Zoning}.
\newblock {\em Journal of the American Planning Association}, 87(2):167--180, 2021.

\bibitem{Manville2022}
Michael Manville, Michael Lens, and Paavo Monkkonen.
\newblock {Zoning and affordability: A reply to Rodr{\'{i}}guez-Pose and Storper}.
\newblock {\em Urban Studies}, 59(1):36--58, 2022.

\bibitem{Shertzer2022}
Allison Shertzer, Tate Twinam, and Randall~P Walsh.
\newblock {Zoning and segregation in urban economic history}.
\newblock {\em Regional Science and Urban Economics}, 94:103652, 2022.

\bibitem{Garde2022}
Ajay Garde and Qi~Song.
\newblock {Housing Affordability Crisis and Inequities of Land Use Change}.
\newblock {\em Journal of the American Planning Association}, 88(1):67--82, 2022.

\bibitem{Talen2003}
Emily Talen and Gerrit Knaap.
\newblock {Legalizing Smart Growth: An Empirical Study of Land Use Regulation in Illinois}.
\newblock {\em Journal of Planning Education and Research}, 22(4):345--359, 2003.

\bibitem{Wickersham2006}
Jay Wickersham.
\newblock {Legal framework: The laws of sprawl and the laws of smart growth}.
\newblock In David~C Soule, editor, {\em Urban sprawl: A comprehensive reference guide}, page Chapter 3. Greenwood Press, Westport, Conn., 2006.

\bibitem{Tyagi2019}
Erika Tyagi and Graham MacDonald.
\newblock {We Need Better Zoning Data. Data Science Can Help. | Urban Institute}, 2019.

\bibitem{Shanks2021}
Brendan Shanks.
\newblock {Land Use Regulations and Housing Development - Evidence from Tax Parcels and Zoning Bylaws in Massachusetts}.
\newblock {\em Working Paper}, 2021.

\bibitem{Song2024}
Jaehee Song.
\newblock {The Effects of Residential Zoning in U.S. Housing Markets}.
\newblock {\em SSRN Electronic Journal}, apr 2024.

\bibitem{Gyourko2011}
Joseph Gyourko, Albert Saiz, and Anita Summers.
\newblock {A New Measure of the Local Regulatory Environment for Housing Markets: The Wharton Residential Land Use Regulatory Index}.
\newblock {\em Urban Studies}, 45(3):693--729, 2008.

\bibitem{mleczko_2023}
Matthew Mleczko and Matthew Desmond.
\newblock {Using natural language processing to construct a National Zoning and Land Use Database}.
\newblock {\em Urban Studies}, 2023.

\bibitem{Ghorbanian2020}
M~Ghorbanian.
\newblock {The Evolution of Urban Zoning from Conventional to Form Based Codes; Introducing Non-Euclidean Zoning Techniques}.
\newblock {\em Int. J. Architect. Eng. Urban Plan}, 30(1):107--118, 2020.

\bibitem{Gray2022}
M.~Nolan Gray.
\newblock {Arbitrary lines : how zoning broke the American city and how to fix it}.
\newblock page 241, 2022.

\bibitem{LiangLinlin2024}
Adam Staveski \& Alex~Horowitz. {Liang, Linlin}.
\newblock {Minneapolis Land Use Reforms Offer a Blueprint for Housing Affordability | The Pew Charitable Trusts}.
\newblock Technical report, Pew Charitable Trust, 2024.

\bibitem{jacobs_1961}
Jane Jacobs.
\newblock {\em {The death and life of great American cities}}.
\newblock Random House, Inc., New York, 1961.

\bibitem{Talen2012}
Emily Talen.
\newblock {City rules : how regulations affect urban form}.
\newblock page 236, 2012.

\bibitem{Garde2017}
Ajay Garde and Cecilia Kim.
\newblock {Form-Based Codes for Zoning Reform to Promote Sustainable Development: Insights From Cities in Southern California}.
\newblock {\em Journal of the American Planning Association}, 83(4):346--364, 2017.

\bibitem{Hansen2014}
Gail Hansen.
\newblock {Design for Healthy Communities: The Potential of Form-Based Codes to Create Walkable Urban Streets}.
\newblock {\em Journal of Urban Design}, 19(2):151--170, 2014.

\bibitem{Chin2024}
Jae~Teuk Chin.
\newblock {Coding for predictive built environments: building and street typology choices in form-based codes}.
\newblock {\em Journal of Asian Architecture and Building Engineering}, 23(1):355--371, 2024.

\bibitem{Hughen2017}
W~Keener Hughen and Dustin~C Read.
\newblock {Analyzing form-based zoning's potential to stimulate mixed-use development in different economic environments}.
\newblock {\em Land Use Policy}, 61:1--11, 2017.

\bibitem{Garde20172}
Ajay Garde and Andrea Hoff.
\newblock {Zoning reform for advancing sustainability: insights from Denver's form-based code}.
\newblock {\em Journal of Urban Design}, 22(6):845--865, 2017.

\bibitem{Zhang2022}
Yingyi Zhang.
\newblock {Evaluating Parametric Form-Based Code for Sustainable Development of Urban Communities and Neighborhoods.}
\newblock {\em International journal of environmental research and public health}, 19(12), jun 2022.

\bibitem{Ameli2015}
Andrea Garfinkel-Castro {S. Hassan Ameli Shima Hamidi} and Reid Ewing.
\newblock {Do Better Urban Design Qualities Lead to More Walking in Salt Lake City, Utah?}
\newblock {\em Journal of Urban Design}, 20(3):393--410, 2015.

\bibitem{Sung2015}
Hyungun Sung, Doohwan Go, Chang-gyu Choi, SangHyun Cheon, and Sungjin Park.
\newblock {Effects of street-level physical environment and zoning on walking activity in Seoul, Korea}.
\newblock {\em Land Use Policy}, 49:152--160, 2015.

\bibitem{Buttazzoni2023}
Adrian Buttazzoni and Leia Minaker.
\newblock {Exploring the relationships between specific urban design features and adolescent mental health: The case of imageability, enclosure, human scale, transparency, and complexity}.
\newblock {\em Landscape and Urban Planning}, 235:104736, 2023.

\bibitem{EnvironmentalProtectionAgency2011}
{Environmental Protection Agency}.
\newblock {Market Acceptance of Smart Growth}.
\newblock Technical report, Washington, D.C.: U.S., 2011.

\bibitem{Talen2021}
Emily Talen.
\newblock {The socio-economic context of form-based codes}.
\newblock {\em Landscape and Urban Planning}, 214:104182, oct 2021.

\bibitem{Park2017}
Yunmi Park.
\newblock {Does new urbanist neighborhood design affect neighborhood turnover?}
\newblock {\em Land Use Policy}, 68:552--562, nov 2017.

\bibitem{Tagtachian2019}
Daniela Tagtachian, Natalie Barefoot, and Adrienne Harreveld.
\newblock {Building by Right: Social Equity Implications of Transitioning to Form-Based Code}.
\newblock {\em J. Affordable Housing \& Commun. Dev. L.}, 28(1), jan 2019.

\bibitem{DenoonStevens2020}
S.~P. Denoon-Stevens and V.~Nel.
\newblock Towards an understanding of proactive upzoning globally and in south africa.
\newblock {\em Land Use Policy}, 97:104708, 2020.

\bibitem{ozay2022}
Erkin {\"{O}}zay.
\newblock {Code as urban vision: a critique of the Buffalo Green Code}.
\newblock {\em Journal of Urban Design}, 27(3):364--385, 2022.

\bibitem{HazelBorys2019}
Emily Talen; Matthew~Lambert {Hazel Borys}.
\newblock {Code Score}, 2019.

\bibitem{Bronin2023}
Sara~C Bronin, Nicholas Abbott, Lindsay Alfano, Steven Della-Giustina, Zoe Merod, and Samuel Terhaar.
\newblock {Zoning by a Thousand Cuts}.
\newblock {\em Pepperdine Law Review 50}, pages 719--784, 2023.

\bibitem{Sahn2021}
A.~Sahn.
\newblock {Racial Diversity and Exclusionary Zoning: Evidence from the Great Migration}.
\newblock 2021.

\bibitem{perez2016formbasedcodes}
Tony Perez.
\newblock Top 10 misconceptions about form-based codes.
\newblock {\em Public Square: A CNU Journal}, May 12 2016.

\bibitem{satterthwaite2016newurbanagenda}
David Satterthwaite.
\newblock Successful, safe and sustainable cities: Towards a new urban agenda.
\newblock {\em Commonwealth Journal of Local Governance}, 19:3--18, 2016.

\bibitem{silva2020localgovernance}
Carlos~Nunes Silva and Anna Trono, editors.
\newblock {\em Local Governance in the New Urban Agenda}.
\newblock Springer, Cham, 2020.

\bibitem{Leyk2018}
Stefan Leyk and Johannes~H Uhl.
\newblock {HISDAC-US, historical settlement data compilation for the conterminous United States over 200 years}.
\newblock {\em Scientific Data}, 5(1):180175, 2018.

\bibitem{Microsoft}
Microsoft.
\newblock {USBuildingFootprints: Computer generated building footprints for the United States}.

\bibitem{Boeing2024}
Geoff Boeing.
\newblock {Modeling and Analyzing Urban Networks and Amenities with OSMnx}.
\newblock 2024.

\end{thebibliography}
\newpage

\appendix

\setcounter{page}{1}
\setcounter{table}{0}
\renewcommand{\thetable}{SI\arabic{table}}
\setcounter{figure}{0}
\renewcommand{\thefigure}{SI\arabic{figure}}
\renewcommand{\thesection}{SI.\arabic{section}}

\section{Supplementary Information: Data}

\section*{Data Availability}
We have provided information on all publicly available data used in our analysis in section \ref{sec:method}. The Landgrid data are confidential and cannot be shared publicly.

\section*{Code Availability}
The code to reproduce these findings are available from the corresponding author upon request.

%\section*{Acknowledgements}

\section*{Author Contributions}
A.S-M. designed research methods. A.S-M. performed data analysis. A.S-M. and E.T wrote the paper.

\section*{Competing Interests}
We have no competing interests to declare.

\section{Supplementary Information}

\begin{table}[!ht]
	\centering
	\caption{\sc{Summary Statistics for Documents}}
	\label{table: document statistics}
	\resizebox{1\textwidth}{!}{
		\begin{tabular}{L{6cm}C{1.5cm}C{1.5cm}C{1.5cm}C{2cm}}\toprule\toprule
			\textbf{Statistic} & \multicolumn{1}{c}{\textbf{Count}} & \multicolumn{1}{c}{\textbf{Mean}} & \multicolumn{1}{c}{\textbf{Median}} & \multicolumn{1}{c}{\textbf{Std Dev}} \\\cmidrule(r){1-5}
			Number of Chunks per Document & 2,723 & 12.46 & 10 & 11.53 \\
			Total Tokens per Document     & 2,723 & 48,975 & 37,740 & 47,253 \\
			Average Tokens per Chunk      & 2,723 & 3,648 & 3,904 & 715 \\
			\\\bottomrule
	\end{tabular}}
	\begin{minipage}{1\linewidth}
		\textsl{Note.---The table reports summary statistics for 2,723 documents used in the regression sample. The columns include the number of observations, the average value, the median, and the standard deviation  for three variables: the number of chunks per document, the total tokens per document, and the average tokens per chunk. Token counts were computed by splitting the text within each document chunk, where tokens are defined as individual words separated by whitespace.}
	\end{minipage}
\end{table}

\begin{figure}[!ht]
    \begin{center}
    \includegraphics[width=\textwidth]{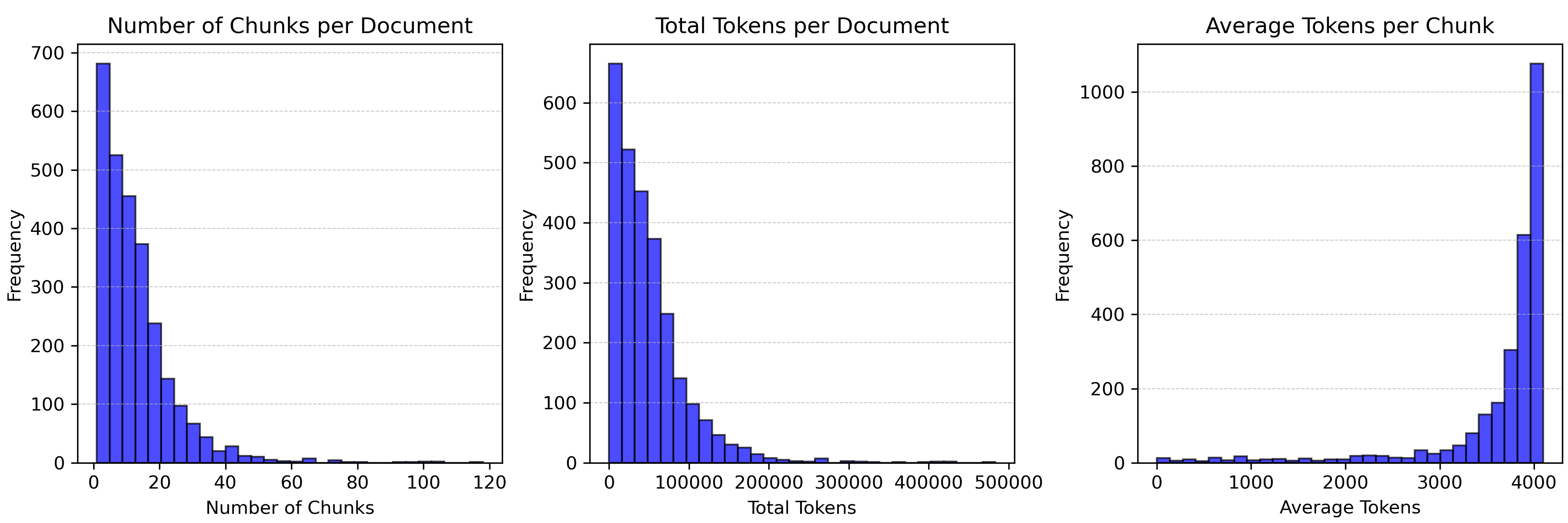}
    \end{center}
    \caption{\textsc{Histograms of Document Metrics.} The figure shows histograms for three document metrics: Number of Chunks per Document, Total Tokens per Document, andAverage Tokens per Chunk.
}
    \label{fig:summary}
\end{figure}

\begin{sidewaystable}[!ht]
    \centering
    \caption{\sc{Summary of Validation Results for Setbacks}}
    \label{table: setbacks}
    \resizebox{1\textwidth}{!}{
    \begin{tabular}{L{2cm}L{4cm}C{2cm}C{2.5cm}C{2cm}L{2.5cm}L{2.5cm}C{2.5cm}C{8cm}C{8cm}L{8cm}}\toprule\toprule
        \textbf{Document Rank} & \textbf{Location} & \textbf{Public Space} & \textbf{Walkability} & \textbf{Urban Form} & \textbf{Human Coder Class} & \textbf{LLM Class} & \textbf{Match} & \textbf{Public Space Description} & \textbf{Walkability Description} & \textbf{Urban Form Description} \\\cmidrule(r){1-11}
        Top & AL, Montgomery & 1 & 0 & 1 & FBC & FBC & TRUE & The lot size, width, depth, shape and orientation, and the minimum building setback lines shall be appropriate for the location of the subdivision and for the type of development and use contemplated. & & Building lines adjacent to streets shall be shown. \\
        Top & FL, LakeHamilton & 1 & 1 & 1 & FBC & FBC & TRUE & Location and dimensions for traffic circulation, designated with arrows, all public and private streets, site access and driveways, pedestrian walks and utility easements within and adjacent to the site. & Location and dimensions for traffic circulation, designated with arrows, all public and private streets, site access and driveways, pedestrian walks and utility easements within and adjacent to the site. & The footprint of all proposed buildings and structures on the site, including setbacks. \\
        Top & IL, Hainesville & 0 & 0 & 0 & Traditional & FBC & FALSE & & & \\
        Top & MO, Columbia & 1 & 1 & 1 & FBC & FBC & TRUE & the area within the street-space between the façade of the building (generally the required building line) and the clear walkway area of the sidewalk & extending beyond the required building line and over the sidewalk or square, open to the street-space & For purposes of form-based zoning standards, the outside corner of a block at the intersection of any two street-spaces \\
        Top & MO, Lee'sSummit & 1 & 1 & 1 & FBC & FBC & TRUE & New buildings shall be aligned with adjacent buildings along the street and conform to established setbacks. & The front setback area shall be designed to maintain a sense of openness of this semi-public space. & Two buildings of the same square footage can have very different 'mass and scale' perceptions, depending on height, setbacks, building materials, and other features. \\
        Top & OH, Rittman & 1 & 0 & 1 & FBC & FBC & TRUE & Where a frontage is divided among the district with different front yard requirements, the deepest front yard shall apply to the entire frontage. & & transition area around water resources that is left in a natural, usually vegetated, state to protect the water resources from runoff pollution.\\
        Top & SC, Mauldin & 1 & 0 & 1 & FBC & FBC & TRUE & Signs must be setback a minimum of five (5) feet from the street right-of-way and shall not obstruct visibility at ingress and egress points onto the site, or other sign structures in the development. & & Where a setback and bufferyard are required along the same property line the requirement with the greatest dimensional width shall be applicable. \\
        Top & TX, Duncanville & 1 & 1 & 1 & FBC & FBC & TRUE & The Streetscape standards shall establish a consistent character along Main Street that relates private development along the street to its public improvements. & Multi-story buildings must also be buffered from nearby single-family areas through the use of setbacks, landscape buffers, and thoroughfares & The Parking setback line shall establish the location behind which surface parking is permitted on each lot. \\
        Top & TX, NassauBay & 1 & 0 & 1 & FBC & FBC & TRUE & none of the above parking areas may be located between the base building line and the pavement edge of a public street. & & The fence, wall or other structure shall not extend beyond the building setback line. \\
        Top & VA, GateCity & 0 & 0 & 0 & Traditional & FBC & FALSE & & & \\
        Bottom & AK, Angoon & 0 & 0 & 0 & Traditional & traditional & TRUE & & & \\
        Bottom & CO, GreenwoodVillage & 1 & 0 & 0 & Traditional & traditional & TRUE & Is properly designed with regard to landscaping, open space, parking, setbacks, and building elevations. & & \\
        Bottom & CO, IdahoSprings & 1 & 0 & 0 & Traditional & traditional & TRUE & That portion of a lot lying between a public street and the nearest parallel front setback line of such lot. & & \\
        Bottom & IL, Elmhurst & 0 & 0 & 0 & Traditional & traditional & TRUE & & & \\
        Bottom & MI, BloomfieldHills & 0 & 0 & 1 & Traditional & traditional & TRUE & & & Setback shall mean the distance required to obtain minimum front, side or rear yard open space provisions. \\
        Bottom & MO, CapeGirardeau & 0 & 0 & 1 & Traditional & traditional & TRUE & & & To regulate and restrict the height, number of stories, size, bulk of buildings and other structures; to restrict the percentage of a lot that may be occupied by buildings or structures; to require setbacks from highways and streets. \\
        Bottom & MO, Odessa & 1 & 0 & 0 & Traditional & traditional & TRUE & All buildings shall be set back from street right-of-way lines to comply with the following front yard requirements & & \\
        Bottom & NC, MintHill & 0 & 1 & 0 & Traditional & traditional & TRUE & & Encroachments: Balconies, stoops, stairs, chimneys, open porches, bay windows, and raised doorways are permitted to encroach into the front setback a maximum of twelve (12) feet. & \\
        Bottom & NM, Clovis & 0 & 0 & 0 & Traditional & traditional & TRUE & & & \\
        Bottom & SC, SurfsideBeach & 0 & 0 & 0 & Traditional & traditional & TRUE & & & \\
        \bottomrule
    \end{tabular}}
    \begin{minipage}{1\linewidth}
        \textsl{Note.—The table summarizes the validation results for zoning codes from ten random documents sampled from the top 20\% and bottom 80\% of the FBC similarity distribution. The analysis focuses on how setbacks relate to form-based design elements. Columns indicate whether paragraphs in the zoning code mention public spaces, walkability, or urban form, based on a rubric. Public Space indicates if setbacks are connected to the design of public spaces or streetscapes. Walkability captures whether setbacks are used to enhance the pedestrian experience. Urban Form identifies whether setbacks are mentioned in the context of urban design or form. Human Coder Class reflects the rubric-based classification assigned by a human evaluator, LLM Class denotes the LLM classification of the zoning code, and Match indicates whether the evaluator’s classification aligns with the LLM classification. Descriptions provide specific examples from the zoning text supporting each rubric.}
    \end{minipage}
\end{sidewaystable}

\begin{sidewaystable}[!ht]
    \centering
    \caption{\sc{Summary of Validation Results for FAR}}
    \label{table: far}
    \resizebox{1\textwidth}{!}{
    \begin{tabular}{L{2cm}L{3cm}C{2cm}C{2.5cm}C{2cm}L{2.5cm}L{2.5cm}C{2.5cm}C{9cm}C{9cm}C{9cm}}\toprule\toprule
        \textbf{Document Rank} & \textbf{Location} & \textbf{Urban Design} & \textbf{Mixed-Use} & \textbf{Context} & \textbf{Human Coder Class} & \textbf{LLM Class} & \textbf{Match} & \textbf{Description Urban Design} & \textbf{Description Mixed-Use} & \textbf{Description Context} \\\cmidrule(r){1-11}
        Top & AL, Montgomery & 0 & 1 & 1 & FBC & FBC & TRUE & & The board of adjustment may…permit the use of a lot or lots under the same ownership in a residential district immediately adjacent to any office, business or industrial district & In specific cases, permit the use of a lot or lots under the same ownership…under such safeguards and conditions as the board may require for the adequate protection of the more restricted property \\
        Top & FL, LakeHamilton & 0 & 0 & 0 & Traditional & FBC & FALSE & & & \\
        Top & IL, Hainesville & 0 & 0 & 0 & Traditional & FBC & FALSE & & & \\
        Top & MO, Columbia & 0 & 1 & 1 & FBC & FBC & TRUE & & Farmer’s Market…Urban Agriculture… Dwelling, live-work… Dwelling, multi-family & Multi-story buildings must also be buffered from nearby single-family areas through the use of setbacks, landscape buffers, and thoroughfares \\
        Top & MO, Lee'sSummit & 1 & 1 & 1 & FBC & FBC & TRUE & Development plan that modifies density, lot size requirements, and design standards. & Primary entrances to ground floor spaces and upper stories shall be oriented to the sidewalk and primary pedestrian ways. & FAR (floor area ratio), density, and lot width standards are outlined in Table 6-2 to adjust intensity based on development needs. \\
        Top & OH, Rittman & 1 & 1 & 1 & FBC & FBC & TRUE & The arrangement and character of all streets shall conform to the Land Use and Thoroughfare Plan. & Planned unit developments shall include common parking areas and ingress/egress points to reduce accidents at intersections with thoroughfares. & The Planning Commission may require sites suitable to the needs created by large-scale developments for schools, parks, and neighborhood purposes. \\
        Top & SC, Mauldin & 1 & 1 & 0 & FBC & FBC & TRUE & The design and arrangement of thoroughfares, civic spaces, and building types & Thoroughfare types can be calibrated and used in creating a walkable community & \\
        Top & TX, Duncanville & 1 & 1 & 1 & FBC & FBC & TRUE & Multi-story buildings must also be buffered from nearby single-family areas through the use of setbacks, landscape buffers, and thoroughfares & In accordance with a comprehensive zoning plan for the purpose of promoting health, safety, morals, and general welfare of the city & Shall include, but may not be limited to: uses, density, lot area, lot width, lot depth, yard depths and widths, building height, building elevations, residential proximity slope, coverage, floor area ratio… \\
        Top & TX, NassauBay & 0 & 1 & 1 & FBC & FBC & TRUE & & To classify, regulate and restrict the use of land, water, buildings and structures…to regulate the intensity of land use & To regulate the height and density of buildings; to regulate the area of yards and other open space about buildings; to regulate the intensity of land use \\
        Top & VA, GateCity & 1 & 0 & 0 & Traditional & FBC & FALSE & Frontage means the minimum width of a lot measured from one side lot line to the other along a straight line on which no point shall be farther away from the street upon which the lot fronts than the building setback line as defined as required herein. & & \\
        Bottom & AK, Angoon & 0 & 0 & 0 & Traditional & Traditional & TRUE & & & \\
        Bottom & CO, GreenwoodVillage & 0 & 0 & 0 & Traditional & Traditional & TRUE & & & \\
        Bottom & CO, IdahoSprings & 0 & 0 & 1 & Traditional & Traditional & TRUE & & & And by other means in accordance with the Comprehensive Plan and with the zoning maps adopted herewith or hereafter \\
        Bottom & IL, Elmhurst & 0 & 0 & 1 & Traditional & Traditional & TRUE & & & Reasonable flexibility is offered through such devices as conditional use, planned development, floor area ratio, and variations. \\
        Bottom & MI, BloomfieldHills & 0 & 0 & 0 & Traditional & Traditional & TRUE & & & \\
        Bottom & MO, CapeGirardeau & 0 & 0 & 0 & Traditional & Traditional & TRUE & & & \\
        Bottom & MO, Odessa & 1 & 0 & 0 & Traditional & Traditional & TRUE & The size of yards and open spaces, density of population and location of buildings. & & \\
        Bottom & NC, MintHill & 1 & 0 & 0 & Traditional & Traditional & TRUE & Landscaping installed within a sight distance shall be set back as far as is practicable from the intersection of the two (2) streets forming the intersection. & & \\
        Bottom & NM, Clovis & 0 & 0 & 0 & Traditional & Traditional & TRUE & & & \\
        Bottom & SC, SurfsideBeach & 1 & 0 & 0 & Traditional & Traditional & TRUE & Convenient, attractive, and harmonious community; to protect and preserve scenic, historic, or ecologically sensitive areas & & \\
        \bottomrule
    \end{tabular}}
    \begin{minipage}{1\linewidth}
        \textsl{Note.—The table summarizes the validation results for zoning codes from ten random documents sampled from the top 20\% and bottom 80\% of the FBC similarity distribution. The analysis evaluates how FAR (floor area ratio) is addressed in zoning documents in relation to form-based design principles. Columns indicate whether paragraphs in the zoning code emphasize urban design over functionality, support for mixed-use or active streetscapes, or contextual regulation of density and intensity instead of uniform standards. Urban Design captures whether FAR is discussed in terms of architectural or urban form. Mixed-Use indicates whether FAR supports active or mixed-use streetscapes. Context reflects whether FAR regulations vary by local development needs or are uniformly applied. Human Coder Class reflects the rubric-based classification assigned by a human evaluator, LLM Class denotes the LLM classification of the zoning code, and Match indicates whether the evaluator’s classification aligns with the LLM classification. Descriptions provide specific examples from the zoning text supporting each rubric.}
    \end{minipage}
\end{sidewaystable}

\begin{table}[!ht]
	\centering
	\caption{\sc{OLS Estimates of the relationship between the similarity to FBC zoning and outcomes}}
	\label{table: urban form continuous}
	\resizebox{1\textwidth}{!}{\begin{tabular}{L{6cm}C{3cm}C{3cm}C{3cm}C{3cm}C{3cm}}\toprule\toprule
			\vspace{0.2cm}
			&(I) &(II)&(III)&(IV)&(V) \\
			\vspace{0.5cm}	
		      \textbf{Setbacks, FARs, plot sizes:} \\
			&\multicolumn{5}{c}{\sc{Panel I. dependent variable: median street setbacks (mts)}}\\\cmidrule(r){2-6}
            Log(FBC zoning similarity)&      -0.419$^{*}$  &      -0.496$^{**}$ &      -0.462$^{**}$ &      -0.499$^{**}$ &      -0.654        \\
            &     (0.231)        &     (0.229)        &     (0.225)        &     (0.229)        &     (0.410)        \\
Observations&        2452        &        2452        &        2452        &        2452        &        2202        \\
R-squared   &        0.13        &        0.13        &        0.16        &        0.16        &        0.09        \\
 
			\vspace{0.2cm}
			&\multicolumn{5}{c}{\sc{Panel II. dependent variable: street setback deviations (mts)}}\\\cmidrule(r){2-6}
            Log(FBC zoning similarity)&      -0.381$^{**}$ &      -0.367$^{**}$ &      -0.335$^{**}$ &      -0.355$^{**}$ &      -0.188        \\
            &     (0.168)        &     (0.167)        &     (0.163)        &     (0.163)        &     (0.193)        \\
Observations&        2450        &        2450        &        2450        &        2450        &        2195        \\
R-squared   &        0.11        &        0.11        &        0.15        &        0.15        &        0.10        \\
 
            \vspace{0.2cm}
			&\multicolumn{5}{c}{\sc{Panel III. dependent variable: log floor-to-area ratio}}\\\cmidrule(r){2-6}
            Log(FBC zoning similarity)&       0.058        &       0.101$^{***}$&       0.099$^{***}$&       0.101$^{***}$&       0.075$^{**}$ \\
            &     (0.039)        &     (0.038)        &     (0.038)        &     (0.038)        &     (0.032)        \\
Observations&        2452        &        2452        &        2452        &        2452        &        2202        \\
R-squared   &        0.11        &        0.16        &        0.17        &        0.17        &        0.13        \\
  
   			\vspace{0.2cm}
      		&\multicolumn{5}{c}{\sc{Panel IV. dependent variable: log minimum plot size}}\\\cmidrule(r){2-6}
            Log(FBC zoning similarity)&      -0.132$^{***}$&      -0.078$^{***}$&      -0.075$^{***}$&      -0.075$^{***}$&      -0.105$^{***}$\\
            &     (0.029)        &     (0.026)        &     (0.026)        &     (0.026)        &     (0.036)        \\
Observations&        2452        &        2452        &        2452        &        2452        &        2202        \\
R-squared   &        0.10        &        0.26        &        0.27        &        0.27        &        0.33        \\
  
   			\vspace{0.5cm}	\\
      
            \textbf{Walkscore, commute, housing:} \\
			&\multicolumn{5}{c}{\sc{Panel V. dependent variable: average walkscore}}\\\cmidrule(r){2-6}
            Log(FBC zoning similarity)&       0.456$^{***}$&       0.394$^{***}$&       0.374$^{***}$&       0.390$^{***}$&       0.416$^{***}$\\
            &     (0.085)        &     (0.084)        &     (0.084)        &     (0.084)        &     (0.105)        \\
Observations&        2687        &        2687        &        2687        &        2687        &        2012        \\
R-squared   &        0.27        &        0.29        &        0.30        &        0.31        &        0.27        \\
 
            \vspace{0.2cm}
            &\multicolumn{5}{c}{\sc{Panel VI. dependent variable: log average commute distance}}\\\cmidrule(r){2-6}
            Log(FBC zoning similarity)&      -0.043$^{***}$&      -0.033$^{***}$&      -0.031$^{***}$&      -0.031$^{***}$&      -0.034$^{**}$ \\
            &     (0.011)        &     (0.011)        &     (0.010)        &     (0.010)        &     (0.013)        \\
Observations&        2588        &        2588        &        2588        &        2588        &        1876        \\
R-squared   &        0.36        &        0.39        &        0.39        &        0.40        &        0.41        \\
 
            \vspace{0.2cm}
            &\multicolumn{5}{c}{\sc{Panel VII. dependent variable: share multi-family housing}}\\\cmidrule(r){2-6}
            Log(FBC zoning similarity)&       0.008        &       0.004        &       0.004        &       0.005        &       0.006        \\
            &     (0.005)        &     (0.005)        &     (0.005)        &     (0.005)        &     (0.007)        \\
Observations&        2721        &        2718        &        2718        &        2718        &        1875        \\
R-squared   &        0.17        &        0.18        &        0.19        &        0.19        &        0.14        \\
             
\textsl{Controls:}\\
State Fixed Effects  &  \checkmark & \checkmark & \checkmark  & \checkmark \\
Location (lat-lon) &  \checkmark & \checkmark & \checkmark & \checkmark  \\
Log Area (km) & & \checkmark & \checkmark & \checkmark  \\
Type of Place & &  & \checkmark & \checkmark   \\
Zoning Vintage Dummies & &  & & \checkmark  \\
\\\bottomrule
	\end{tabular}}
	\begin{minipage}{1\linewidth}											
		\scriptsize \textsl{Note.---The Table reports estimates from regressing urban form outcomes on the continuous measure of FBC adoption. The unit of observation is the census place. Panels I through IV report estimates for median street setbacks, street setback deviations, floor-to-area ratio, and minimum plots size, respectively.  Panels V through VII report estimates for walkscore, commute distance, and the share of multi-family housing, respectively. Column I includes state fixed effects, Column II adds location controls (latitude-longitude), and column III further controls for log area (km²) and type of place (borough, city, town, village). Column IV incorporates zoning vintage dummies (1982-1996, 1996-2008, 2008-2016, 2016-2021). Column V reports estimates for outcomes computed for neighborhoods developed after 1950 in each place.}								
	\end{minipage}	
\end{table}

\begin{table}[!ht]
	\centering
	\caption{\sc{OLS Estimates of the relationship between the similarity to FBC zoning and outcomes}}
	\label{table: urban form discrete}
	\resizebox{1\textwidth}{!}{\begin{tabular}{L{6cm}C{3cm}C{3cm}C{3cm}C{3cm}C{3cm}}\toprule\toprule
			\vspace{0.2cm}
			&(I) &(II)&(III)&(IV)&(V) \\
			\vspace{0.5cm}	
		      \textbf{Setbacks, FARs, plot sizes:} \\
			&\multicolumn{5}{c}{\sc{Panel I. dependent variable: median street setbacks (mts)}}\\\cmidrule(r){2-6}
            High-FBC (top 20\%)&      -0.719$^{**}$ &      -0.849$^{***}$&      -0.794$^{***}$&      -0.842$^{***}$&      -1.360$^{***}$\\
            &     (0.283)        &     (0.276)        &     (0.273)        &     (0.279)        &     (0.346)        \\
Observations&        2452        &        2452        &        2452        &        2452        &        2202        \\
R-squared   &        0.13        &        0.13        &        0.16        &        0.16        &        0.09        \\
 
			\vspace{0.2cm}
			&\multicolumn{5}{c}{\sc{Panel II. dependent variable: street setback deviations (mts)}}\\\cmidrule(r){2-6}
            High-FBC (top 20\%)&      -0.657$^{***}$&      -0.636$^{***}$&      -0.578$^{***}$&      -0.607$^{***}$&      -0.500$^{*}$  \\
            &     (0.196)        &     (0.195)        &     (0.190)        &     (0.192)        &     (0.271)        \\
Observations&        2450        &        2450        &        2450        &        2450        &        2195        \\
R-squared   &        0.11        &        0.11        &        0.15        &        0.15        &        0.10        \\
 
            \vspace{0.2cm}
			&\multicolumn{5}{c}{\sc{Panel III. dependent variable: log floor-to-area ratio}}\\\cmidrule(r){2-6}
            High-FBC (top 20\%)&       0.096$^{*}$  &       0.166$^{***}$&       0.158$^{***}$&       0.162$^{***}$&       0.152$^{***}$\\
            &     (0.049)        &     (0.048)        &     (0.047)        &     (0.048)        &     (0.042)        \\
Observations&        2452        &        2452        &        2452        &        2452        &        2202        \\
R-squared   &        0.11        &        0.16        &        0.17        &        0.17        &        0.14        \\
  
   			\vspace{0.2cm}
      		&\multicolumn{5}{c}{\sc{Panel IV. dependent variable: log minimum plot size}}\\\cmidrule(r){2-6}
            High-FBC (top 20\%)&      -0.204$^{***}$&      -0.117$^{***}$&      -0.112$^{***}$&      -0.114$^{***}$&      -0.137$^{***}$\\
            &     (0.039)        &     (0.036)        &     (0.036)        &     (0.036)        &     (0.047)        \\
Observations&        2452        &        2452        &        2452        &        2452        &        2202        \\
R-squared   &        0.10        &        0.26        &        0.27        &        0.27        &        0.33        \\
  
   			\vspace{0.5cm}	\\
      
            \textbf{Walkscore, commute, housing:} \\
			&\multicolumn{5}{c}{\sc{Panel IV. dependent variable: average walkscore}}\\\cmidrule(r){2-6}
            High-FBC (top 20\%)&       0.420$^{***}$&       0.317$^{***}$&       0.303$^{***}$&       0.327$^{***}$&       0.352$^{**}$ \\
            &     (0.116)        &     (0.116)        &     (0.116)        &     (0.116)        &     (0.145)        \\
Observations&        2687        &        2687        &        2687        &        2687        &        2012        \\
R-squared   &        0.27        &        0.29        &        0.30        &        0.30        &        0.26        \\
 
            \vspace{0.2cm}
            &\multicolumn{5}{c}{\sc{Panel V. dependent variable: log average commute distance}}\\\cmidrule(r){2-6}
            High-FBC (top 20\%)&      -0.052$^{***}$&      -0.036$^{***}$&      -0.033$^{**}$ &      -0.033$^{**}$ &      -0.027$^{*}$  \\
            &     (0.014)        &     (0.013)        &     (0.013)        &     (0.013)        &     (0.017)        \\
Observations&        2588        &        2588        &        2588        &        2588        &        1876        \\
R-squared   &        0.36        &        0.39        &        0.39        &        0.39        &        0.41        \\
 
            \vspace{0.2cm}
            &\multicolumn{5}{c}{\sc{Panel VI. dependent variable: share multi-family housing}}\\\cmidrule(r){2-6}
            High-FBC (top 20\%)&       0.020$^{***}$&       0.014$^{**}$ &       0.014$^{**}$ &       0.015$^{**}$ &       0.011        \\
            &     (0.007)        &     (0.007)        &     (0.007)        &     (0.007)        &     (0.010)        \\
Observations&        2721        &        2718        &        2718        &        2718        &        1875        \\
R-squared   &        0.17        &        0.18        &        0.19        &        0.19        &        0.14        \\

\textsl{Controls:}\\
State Fixed Effects  &  \checkmark & \checkmark & \checkmark  & \checkmark \\
Location (lat-lon) &  \checkmark & \checkmark & \checkmark & \checkmark  \\
Log Area (km) & & \checkmark & \checkmark & \checkmark  \\
Type of Place & &  & \checkmark & \checkmark   \\
Zoning Vintage Dummies & &  & & \checkmark  \\
\\\bottomrule
	\end{tabular}}
	\begin{minipage}{1\linewidth}											
		\scriptsize \textsl{Note.---The Table reports estimates from regressing urban form outcomes on high-FBC
(top 20\% of the FBC distribution). The unit of observation is the census place. Panels I through IV report estimates for median street setbacks, street setback deviations, floor-to-area ratio, and minimum plots size, respectively.  Panels V through VII report estimates for walkscore, commute distance, and the share of multi-family housing, respectively. Column I includes state fixed effects, Column II adds location controls (latitude-longitude), and column III further controls for log area (km²) and type of place (borough, city, town, village). Column IV incorporates zoning vintage dummies (1982-1996, 1996-2008, 2008-2016, 2016-2021). Column V reports estimates for outcomes computed for neighborhoods developed after 1950 in each place.}								
	\end{minipage}	
\end{table}	
\end{document}